\documentclass[12pt]{article}
\usepackage{amsmath}
\usepackage{amsfonts}
\usepackage{graphicx}

\newtheorem{theorem}{Theorem}[section]
\newtheorem{lemma}[theorem]{Lemma}
\newtheorem{proposition}[theorem]{Proposition}

\newtheorem{Remark}[theorem]{Remark}
\newenvironment{remark}{\begin{Remark}\rm}{\end{Remark}}

\newenvironment{proof}
    {\rm \trivlist \item[\hskip \labelsep{\bf Proof. }]}
    {\hspace*{\fill}$\Box$\endtrivlist}
\newenvironment{varproof}
    {\rm \trivlist \item[\hskip \labelsep{\bf Proof}]}
    {\hspace*{\fill}$\Box$\endtrivlist}
\numberwithin{equation}{section}

\DeclareMathOperator*{\Tr}{Tr}
\DeclareMathOperator*{\Ai}{Ai}
\DeclareMathOperator*{\supp}{supp}

\DeclareMathOperator*{\RE}{Re}
\DeclareMathOperator*{\IM}{Im}
\renewcommand{\Re}{\RE}
\renewcommand{\Im}{\IM}
\newcommand{\ds}{\displaystyle}

\begin{document}
\title{Universality of the double scaling limit in
random matrix models}
\author{Tom Claeys and Arno B.J. Kuijlaars}

\maketitle
\begin{abstract}
We study unitary random matrix ensembles in the critical case
where the limiting mean eigenvalue density vanishes quadratically
at an interior point of the support. We establish universality of
the limits of the eigenvalue correlation kernel at such a critical
point in a double scaling limit. The limiting kernels are constructed
out of functions associated with the second Painlev\'e equation.
This extends a result of Bleher and Its for the
special case of a critical quartic potential.

The two main tools we use are equilibrium measures and
Riemann-Hilbert problems.  In our treatment of equilibrium measures
we allow a negative density near the critical point, which
enables us to treat all cases simultaneously. The asymptotic
analysis of the Riemann-Hilbert problem is done with the
Deift/Zhou steepest descent analysis. For the construction of
a local parametrix at the critical point we introduce a
modification of the approach of Baik, Deift, and Johansson
so that we are able to satisfy the required jump properties
exactly.
\end{abstract}
\section{Introduction}
We consider the unitary random matrix model
\begin{equation}
 Z_{n,N}^{-1}\exp(-N\Tr V(M))dM \label{1-randommatrixmodel}
\end{equation}
defined on Hermitian $n \times n$ matrices $M$ in a critical
regime where the limiting mean density of eigenvalues vanishes at
an interior point. It is a basic fact of random matrix theory
\cite{Deift,Mehta} that the eigenvalues of the
random matrix ensemble (\ref{1-randommatrixmodel}) follow a
determinantal point process with correlation kernel
\begin{equation}
    K_{n,N}(x,y)=e^{-\frac{N}{2}V(x)}e^{-\frac{N}{2}V(y)}
    \sum_{k=0}^{n-1}p_{k,N}(x)p_{k,N}(y),
\label{1-kern}
\end{equation}
where $p_{k,N}$ denotes the $k$th degree orthonormal polynomial with
respect to the weight $e^{-NV(x)}$ on $\mathbb R$.

We assume in this paper that the confining potential
$V : \mathbb R \to \mathbb R$ in
(\ref{1-randommatrixmodel}) is real analytic and that it satisfies
the growth condition
\begin{equation}
    \frac{V(x)}{\log (x^2+1)} \to +\infty \quad\textrm{ as }
    |x|\to +\infty. \label{1-voorwaardeV}
\end{equation}
These assumptions ensure that the mean eigenvalue density
$\frac{1}{n}K_{n,N}(x,x)$ has a limit as $n, N \to \infty$,
$n/N \to 1$, see e.g.\ \cite{Deift},
which we denote by $\psi_V(x)$.
It is known that $\psi_V$ is the density of the measure
$\mu_V$ which minimizes the weighted energy
\begin{equation} \label{1-minenergie}
I_V(\mu)=\iint \log \frac{1}{|s-t|}d\mu(s)d\mu(t)
    + \int V(t) d\mu(t)
\end{equation}
among all probability measure on $\mathbb R$. The measure
$\mu_V$ is called the equilibrium measure in the external
field $V$.
The fact that $V$ is real analytic ensures that the support
$S_V = \supp(\mu_V)$ consists of a finite union
of intervals \cite{DKM}.

It is a remarkable fact that local scaling limits of the kernel (\ref{1-kern})
depend only on the nature of the density $\psi_V$.
This has been proved rigorously in the bulk of the spectrum
for a quartic $V$ in \cite{BI1} and for general real analytic $V$
in \cite{DKMVZ2}. Indeed, if $\psi_V(x^*) > 0$,
then
\begin{equation} \label{bulk-scaling}
    \lim_{n \to \infty}
    \frac{1}{n \psi_V(x^*)} K_{n,n}\left(x^* + \frac{u}{n\psi_V(x^*)},
    x^* + \frac{v}{n \psi_V(x^*)}\right)
    = K^{bulk}(u,v)
    \end{equation}
exists, and
\begin{equation} \label{sine-kernel}
    K^{bulk}(u,v) = \frac{\sin \pi(u-v)}{\pi (u-v)}.
\end{equation}
The scaling limits are different at special points of the spectrum.
At edge points of the spectrum the density $\psi_V$ typically vanishes like a square
root, and then it is known that for some constant $c > 0$,
\begin{equation} \label{edge-scaling}
    \lim_{n \to \infty}
    \frac{1}{(cn)^{2/3}}
    K_{n,n}\left(x^* + \frac{u}{(cn)^{2/3}}, x^*+\frac{v}{(cn)^{2/3}}\right)
    = K^{edge}(u,v)
    \end{equation}
where
\begin{equation} \label{Airy-kernel}
    K^{edge}(u,v)  = \frac{\Ai(u)\Ai'(v) - \Ai'(u)\Ai(v)}{u-v}
\end{equation}
and $\Ai$ is the Airy function. The Airy kernel is related to
the Tracy-Widom distribution \cite{TW}.
In (\ref{edge-scaling}) we have assumed that $x^*$
is a right edge point. For a left edge point we change $u \mapsto -u$, $v \mapsto -v$
in the left-hand side of (\ref{edge-scaling}).

Other special points in the spectrum include
\begin{itemize}
\item Edge points where the density vanishes to a higher
order. The possible edge point behaviors
(at a right end point $x^*$) are
\begin{equation} \label{edge-behavior}
    \psi_V(x) = c (x^*-x)^{2k + \frac{1}{2}} (1+o(1))
    \quad \mbox{ as } x \to x^*+
\end{equation}
where $c > 0$ and $k$ is a non-negative integer.
\item Interior points where the density vanishes. Then
\begin{equation} \label{interior-vanishing}
    \psi_V(x) = c (x- x^*)^{2k} ( 1 +o(1))
    \quad \mbox{ as } x \to x^*
\end{equation}
where $c > 0$ and $k$ is a positive integer.
\end{itemize}
In these critical cases it is believed that the local
scaling limit at $x^*$ of the kernel only depends on the order of
vanishing of the density at $x^*$ \cite{BE}.

The case where $\psi_V$ vanishes
quadratically at an interior point of $S_V$, that
is, the case $k=1$ in (\ref{interior-vanishing}),
was considered by Bleher and Its \cite{BI2}
for the case of a critical quartic potential
$V(x) =  \frac{g}{4} x^4 + \frac{t}{2} x^2$,
with $g > 0$ and $t = t_c = -2\sqrt{g}$. Then
\[ \psi_V(x) = \frac{1}{2\pi} gx^2 \sqrt{\frac{4}{\sqrt{g}}-x^2},
    \qquad \mbox{for } x \in [-2 g^{-1/4}, 2 g^{-1/4}], \]
so that $\psi_V$ vanishes quadratically at the origin.
Bleher and Its consider the double scaling limit
where  $t$ changes with $n$ and tends to $t_c$
as $n \to \infty$ in such a way that $n^{2/3}(t-t_c)$
remains constant. For the quartic potential this
is equivalent to considering (\ref{1-randommatrixmodel})
where $n, N \to \infty$, $n/N \to 1$, such that
\begin{equation} \label{nN-to-infty}
    \lim_{n, N \to \infty}
    n^{2/3} \left( \frac{n}{N}-1\right)
    \end{equation}
exists. Bleher and Its gave a one-paremeter family
$K^{crit}(u,v;s)$ of limiting kernels, depending on $s \in \mathbb R$,
so that for some $c > 0$,
\begin{equation}
\lim_{n,N \to \infty} \frac{1}{(c n)^{1/3}}
 K_{n,N}\left(\frac{u}{(cn)^{1/3}}, \frac{v}{(cn)^{1/3}}\right)
        \label{1-Phikernel}
        = K^{crit}(u,v;s)
\end{equation}
where $s$ is proportional to the value of the limit (\ref{nN-to-infty}).

The critical kernels are expressed in terms of so-called $\psi$-functions
associated with the Hastings-McLeod solution of the Painlev\'e II equation
\cite{HM}.
Consider as in \cite{BI2} the linear differential equations
for a $2$-vector (or $2 \times 2$ matrix) $\Psi = \Psi(\zeta;s)$,
\begin{equation} \label{Psi-diffeqns}
    \frac{d}{d\zeta} \Psi = A \Psi, \qquad
    \frac{\partial}{\partial s} \Psi = B \Psi
    \end{equation}
where
\begin{equation} \label{Psi-diffA}
    A = A(\zeta;s)  =
    \begin{pmatrix} 4 \zeta q & 4 \zeta^2 + s + 2q^2 + 2 r \\
    -4 \zeta^2 -s - 2q^2 + 2r & - 4 \zeta q \end{pmatrix},
    \end{equation}
and
\begin{equation} \label{Psi-diffB}
    B = B(\zeta;s)  =
    \begin{pmatrix} q & \zeta \\ - \zeta & -q
    \end{pmatrix}.
    \end{equation}
The compatibility condition for (\ref{Psi-diffeqns})
is that $q = q(s)$ satisfies the Painlev\'e II equation
$q'' = sq + 2q^3$ and that $r = r(s) = q'(s)$.
We assume that $q(s)$ is the Hastings-McLeod solution
of Painlev\'e II, which is characterized by the
asymptotic condition
\[ q(s) = \Ai(s)(1+o(1)) \qquad \mbox{ as } s \to +\infty. \]
The critical kernels op \cite{BI2} are given by
\begin{equation} \label{critK1}
    K^{crit}(u,v;s) =
    \frac{\Phi^1(u;s) \Phi^2(v;s) - \Phi^2(u;s) \Phi^1(v;s)}
    {\pi (u-v)}
    \end{equation}
where $ \begin{pmatrix} \Phi^1(\zeta;s) \\ \Phi^2(\zeta;s) \end{pmatrix}$
is the special solution to (\ref{Psi-diffeqns})
which is real for real $\zeta$, satisfies
\[ \Phi^1(-\zeta;s) = \Phi^1(\zeta;s),
    \qquad \Phi^2(-\zeta;s) = -\Phi^2(\zeta;s) \]
and has  asymptotics on the real line
\[ \begin{aligned}
 \Phi^1(\zeta;s) & = \cos \left( \frac{4}{3} \zeta^3 + s \zeta\right)
    + O(\zeta^{-1}), \\
    \Phi^2(\zeta;s) & = -\sin \left(\frac{4}{3} \zeta^3 + s \zeta\right)
    + O(\zeta^{-1})
    \end{aligned}
    \quad \mbox{ as } \zeta \to \pm \infty.
\]

If we put
\begin{equation} \label{Phi1and2}
    \Phi_1 = \Phi^1 + i \Phi^2,
    \qquad \Phi_2 = \Phi^1 - i \Phi^2
    \end{equation}
then
\begin{equation} \label{critK2}
    K^{crit}(u,v;s) =
    \frac{-\Phi_1(u;s) \Phi_2(v;s) + \Phi_2(u;s) \Phi_1(v;s)}
    {2\pi i (u-v)}
    \end{equation}
and $\begin{pmatrix} \Phi_1 \\ \Phi_2 \end{pmatrix}$ is a special
solution of the differential equations
\begin{equation}
   \label{1-diiffvglpsi1}
\frac{d}{d\zeta} \Psi(\zeta;s)
    = \begin{pmatrix}
     -4i\zeta^2-i(s+2q^2)  & 4\zeta q+2ir  \\
     4\zeta q -2i r   & 4i\zeta^2+i(s+2q)
    \end{pmatrix}
    \Psi(\zeta;s)
\end{equation}
and
\begin{equation}
    \label{1-diiffvglpsi2}
\frac{\partial}{\partial s}
    \Psi(\zeta;s)
     =  \begin{pmatrix}
     - i\zeta & q \\ q & i\zeta \end{pmatrix}
    \Psi(\zeta;s).
\end{equation}
The equations (\ref{1-diiffvglpsi1})-(\ref{1-diiffvglpsi2}) for the $\psi$-functions
correspond to the ones used by Flaschka and Newell \cite{FN} and we will
also use those in what follows.
The vector $\begin{pmatrix} \Phi_1 \\ \Phi_2 \end{pmatrix}$
is the unique solution of (\ref{1-diiffvglpsi1}) with
asymptotics
\begin{equation}\label{1-asymptotiekphi}
    e^{i(\frac{4}{3} \zeta^3 + s \zeta)}
    \begin{pmatrix}
    \Phi_1(\zeta; s) \\
    \Phi_2(\zeta;s)
    \end{pmatrix}
     =
 \begin{pmatrix}  1 \\ 0 \end{pmatrix}
    +O\left(\zeta^{-1}\right)
    \end{equation}
as $\zeta \to \infty$ uniformly in
$\varepsilon \leq \arg\zeta \leq \pi - \varepsilon$
for any $\varepsilon > 0$.

Before discussing our results, we like to point out
an integral formula for
the kernel $K^{crit}(u,v;s)$. If we take a derivative
of (\ref{critK1}) with respect to $s$ and use
(\ref{Psi-diffeqns}) and (\ref{Psi-diffB}), we get
after some calculations
\[ \frac{d}{ds} K^{crit}(u,v;s) =
    \frac{1}{\pi}
    \left[ \Phi^1(u;s) \Phi^1(v;s) + \Phi^2(u;s) \Phi^2(v;s) \right].
\]
Using the Deift/Zhou steepest-descent method for $s\to -\infty$ as
done in \cite{DZ2}, one can show that $K^{crit}(u,v;s) \to 0$ as
$s \to -\infty$, so that we get
\begin{equation} \label{critK3}
    K^{crit}(u,v;s)
    = \frac{1}{\pi}
        \int_{-\infty}^s
  \left[ \Phi^1(u;\sigma) \Phi^1(v;\sigma) + \Phi^2(u;\sigma) \Phi^2(v;\sigma) \right]
  d\sigma.
\end{equation}
Since $\Phi^1(\zeta;s)$ and $\Phi^2(\zeta;s)$
are real for real $\zeta$, formula (\ref{critK3})
clearly shows that $K^{crit}(u,u;s) > 0$, as it
should be.

\section{Statement of results}

It is the aim of this paper to show that the kernel
$K^{crit}(u,v;s)$ is a universal limit.
Whenever the limiting mean eigenvalue density
$\psi_V$ vanishes quadratically at an interior point,
the correlation kernel $K_{n,N}$ has a double scaling
limit given by (\ref{1-Phikernel}).

In our Theorem \ref{theorem-1-maintheorem} below, we use
the equilibrium measure  $\omega_{S}$ of a compact set $S \subset \mathbb R$.
This is the unique probability measure on $S$ that minimizes the logarithmic energy
\begin{equation}
I(\mu)=\iint \log \frac{1}{|s-t|}d\mu(s)d\mu(t)
\end{equation}
among all Borel probability measures $\mu$ on $S$. If $S$
is a single interval $[a,b]$, then $\omega_S$ has a density
$w_S$ given by
\[ w_S(x) = \frac{1}{\pi \sqrt{(b-x)(x-a)}},
    \qquad x \in (a,b). \]
If $S$ is a finite union of disjoint intervals, say
$S = \bigcup_{j=1}^n [a_j, b_j]$ with $a_j < b_j < a_{j+1}$. Then
$\omega_S$ has density
\begin{equation} \label{wSdensity}
    w_S(x) =
    \frac{|p(x)|}{\pi  \sqrt{\prod_{j=1}^n|(b_j-x)(x-a_j)|}},
    \qquad x \in \bigcup_{j=1}^n (a_j, b_j),
\end{equation}
where $p(x)$ is a monic polynomial of degree $n-1$ with exactly
one zero in each of the gaps $(b_j, a_{j+1})$, $j=1, \ldots, n-1$,
see e.g.\ \cite[Lemma 4.4.1]{StTo}.
Note that (\ref{wSdensity}) has an extension to an analytic function
in $\mathbb C \setminus (\mathbb R \setminus S^o)$, where
$S^o = \bigcup_j (a_j,b_j)$,
which is a fact that we will use in what follows.

The following is our main result.

\begin{theorem}\label{theorem-1-maintheorem}
Let $V$ be real analytic on $\mathbb R$ such
that $\lim\limits_{x \to \pm \infty} \frac{V(x)}{\log(x^2+1)} = +\infty$.
Let $\psi_V$ be the density of the equilibrium measure
in the external field, and let $x^*$ be an interior
point of $S_V = \supp(\psi_V)$ which is  such that
\[ \psi_V(x^*) = \psi_V'(x^*) = 0, \qquad \psi_V''(x^*) > 0. \]
Let $n, N \to \infty$ such that the limit
\[ \lim_{n,N \to \infty} n^{2/3} \left(\frac{n}{N} - 1\right)
 = L \]
exists with $L \in \mathbb R$. Let $K_{n,N}$ be the
correlation kernel {\rm (\ref{1-kern})} for the eigenvalues of
the random matrix model {\rm (\ref{1-randommatrixmodel})}.
Then there exist constants $c > 0$ and
$s \in \mathbb R$ such that
\begin{equation} \label{main1}
 \lim_{n,N \to \infty} \frac{1}{(c n)^{1/3}}
        K_{n,N}\left(x^*+\frac{u}{(cn)^{1/3}}, x^*+\frac{v}{(cn)^{1/3}}\right)
    = K^{crit}(u,v;s)
\end{equation}
uniformly for $u, v$ in compact subsets of $\mathbb R$.

Explicit formulas for the constants $c$ and $s$ are
\begin{equation} \label{main2}
    c = \frac{\pi \psi_V''(x^*)}{8}
    \end{equation}
and
\begin{equation} \label{main3}
    s = L \frac{\pi}{c^{1/3}} w_{S_V}(x^*),
    \end{equation}
where $w_{S_V}$ is the density of the equilibrium measure of
$S_V$.
\end{theorem}
As noted before, Bleher and Its \cite{BI2} proved (\ref{main1})
for the case of a critical quartic $V$. See  \cite{BI3}
for a rigorous expansion of the free energy in this critical
case.

\begin{remark}
The random matrix model (\ref{1-randommatrixmodel}) may
be generalized to include a spectral singularity at the origin
\begin{equation} \label{spec-sing}
    Z_{n,N}^{-1} |\det M|^{2\alpha} \exp(-N \Tr V(M)) dM,
    \qquad \alpha > -1/2.
    \end{equation}
If $\psi_V(0) > 0$ and $n = N \to \infty$, then the scaled
limit of the correlation kernels is a Bessel kernel which involves
Bessel functions of order $\alpha \pm \frac{1}{2}$, see
\cite{ADMN,KV}. In the multicritical case where $\psi_V$
vanishes quadratically at $0$, an analogue of Theorem 2.1
is valid. In work in progress \cite{CKV}, we are considering the
double scaling limit of (\ref{spec-sing}) and we show
that the limiting kernels are expressed in terms of the
$\psi$-functions associated with a special solution
of the general Painlev\'e II equation
\[ q'' = sq + 2 q^3 - \alpha. \]
\end{remark}

\bigskip

The main ingredients in the proof of Theorem \ref{theorem-1-maintheorem} are
equilibrium measures and Riemann-Hilbert problems. We give some
comments on both.

\paragraph{Equilibrium measures.}
Recall that the equilibrium measure in external field $V$
minimizes (\ref{1-minenergie}).
We need to know how the equilibrium measure $\mu_V$ in the
external field changes as a result of a change in $V$.
The particular modification we consider here is
\[ V_t = \frac{1}{t} V, \qquad t > 0, \]
so that $V_1 = V$. Let us put
\[ \mu_t = \mu_{V_t}. \]
Then it is known that $t\mu_t$ and $S_t$ are increasing
as a function of $t$, see e.g.\ \cite{DK,SaTo,Totik}.
We also have the Buyarov-Rakhmanov formula \cite{BR}
\begin{equation} \label{1-BuRaformule}
    \mu_t = \frac{1}{t} \int_0^{t} \omega_{\supp(\mu_{\tau})} d \tau,
\end{equation}
which expresses the equilibrium measure in the external field
as an average of equilibrium measures of sets. A
consequence of (\ref{1-BuRaformule}) is that
\begin{equation} \label{1-BuRa2}
    \left. \frac{d }{dt} \left( t\mu_t\right) \right|_{t = 1} = \omega_{S_V},
    \end{equation}
which partly explains why the equilibrium measure
$\omega_{S_V}$ plays a role in the formula (\ref{main3})
for $s$.

\begin{figure}[t]
\begin{center}
\includegraphics[scale=0.18,angle=270]{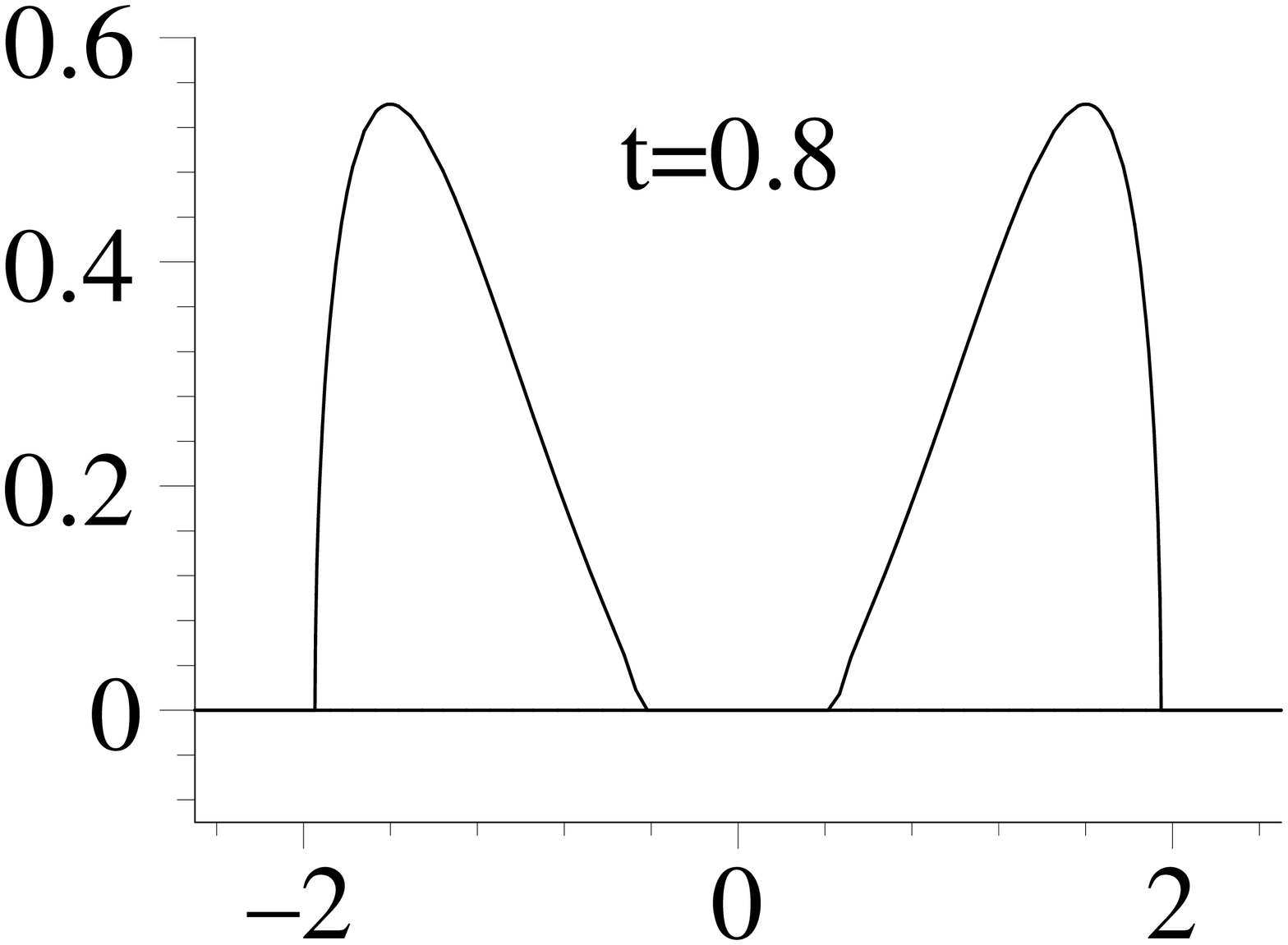}
\includegraphics[scale=0.18,angle=270]{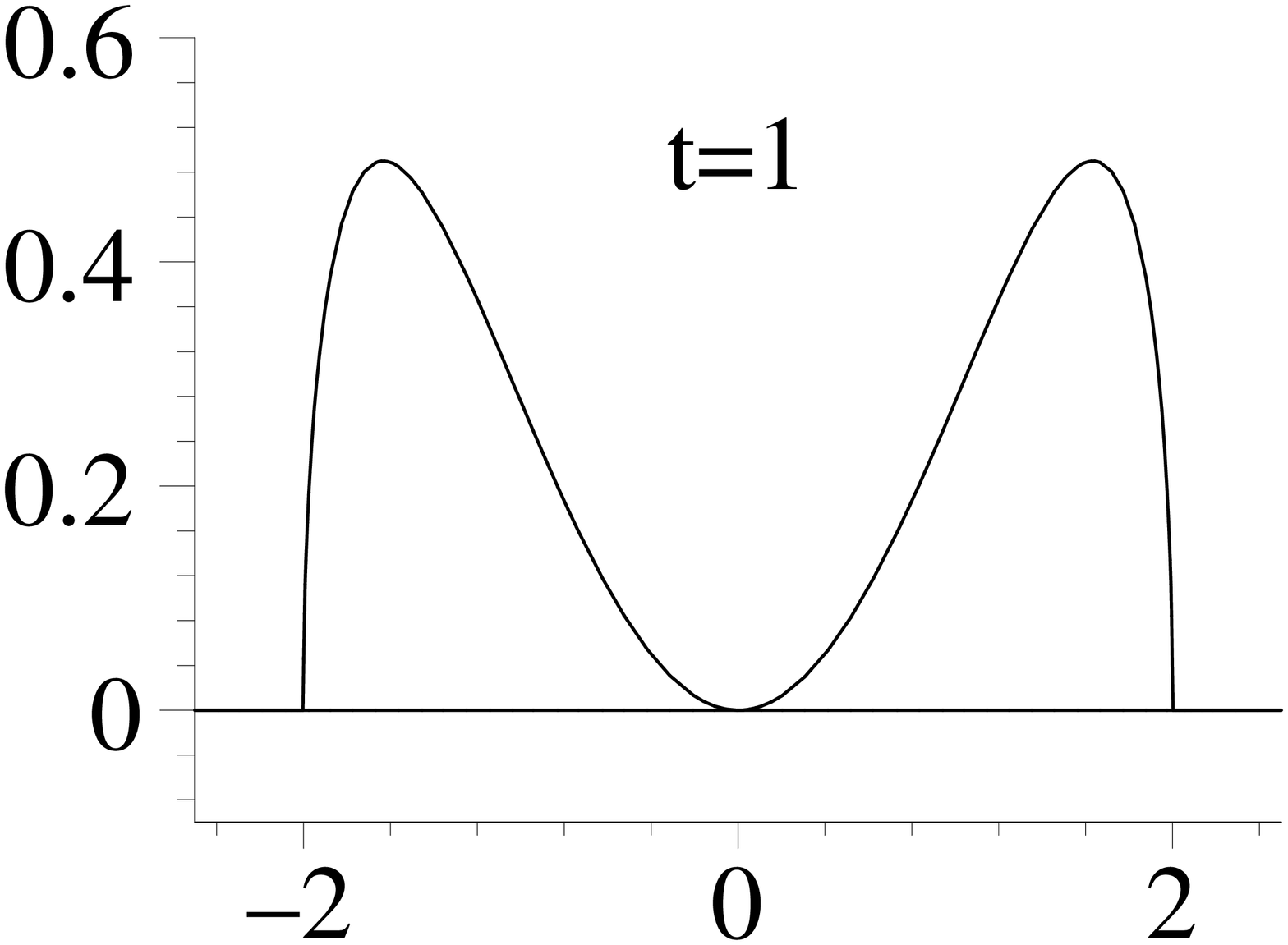}
\includegraphics[scale=0.18,angle=270]{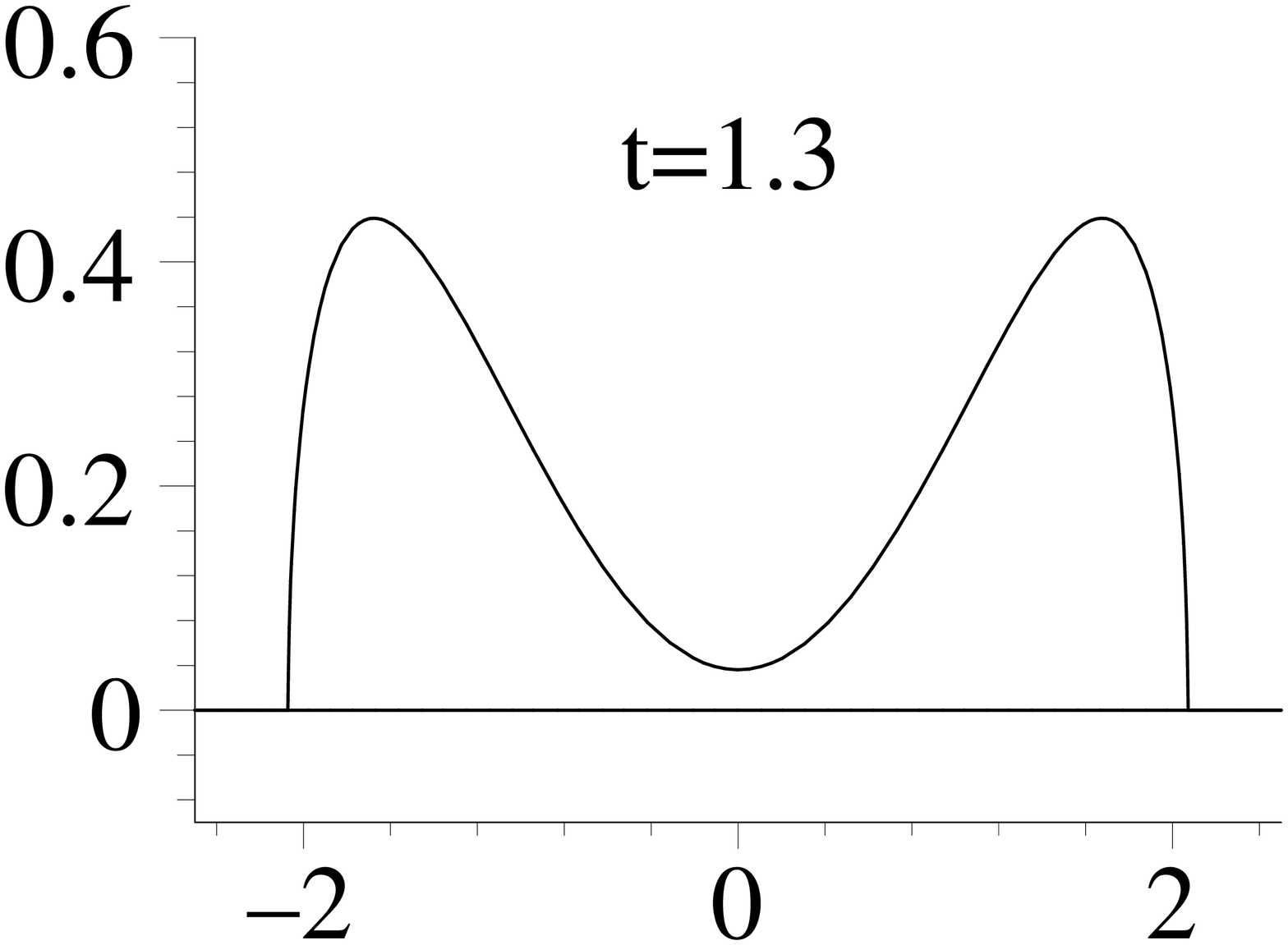}
\end{center}
\caption{The density of $\mu_t$ for $V(x)=\frac{x^4}{4}-x^2$
with $t$ equal to $0.8$, $1$ and $1.3$, respectively.} \label{figure-tweedeschaling}
\end{figure}

For the case of interest in this paper we have that $\psi_V$
vanishes at $x = x^*$. Then for $t > 1$, there is a positive
density at $x^*$, while for $t < 1$, $x^*$ is out of the support
of $\mu_t$. For $t$ slightly less than $1$, there is a gap in
$\supp(\mu_t)$, see Figure \ref{figure-tweedeschaling}. An
asymptotic analysis based on the equilibrium measure $\mu_t$ would
require a discussion of the two different situations $t > 1$ and
$t < 1$, as is done in \cite{BDJ}.

Therefore we found it convenient to introduce a modification
of the equilibrium problem in external field, which will
enable us to treat both cases simultaneously. The modification
we make is that we do not require the measure to be non-negative in a neighborhood of the
point $x^*$. For a sufficiently small $\delta_0 > 0$, we consider the
problem to minimize
\begin{equation} \label{mod-energy1}
    I_{V_t}(\nu) = \iint \log \frac{1}{|x-y|}d\nu(x)d\nu(y)
    + \frac{1}{t} \int V(x) d\nu(x)
\end{equation}
among all signed measures $\nu = \nu^+ - \nu^-$ on $\mathbb R$,
where $\nu^{\pm}$ are nonnegative measures, such that
\begin{equation} \label{mod-energy2}
    \int d\nu = 1, \quad \mbox{ and }
    \quad \supp(\nu^-) \subset [x^*-\delta_0, x^*+\delta_0].
\end{equation}
We denote the minimizer by $\nu_t$ and we let $S_t = \supp(\nu_t)$.

\begin{figure}[t]
\begin{center}
\includegraphics[scale=0.18,angle=270]{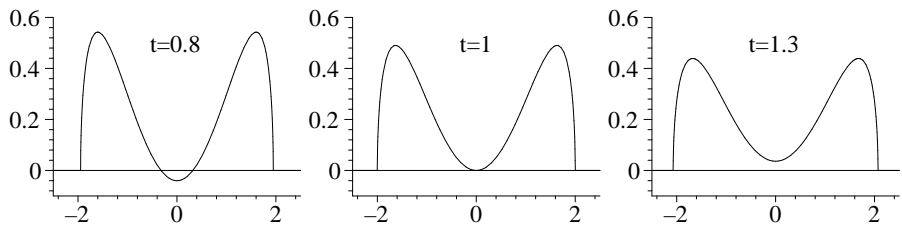}
\includegraphics[scale=0.18,angle=270]{nu1.eps}
\includegraphics[scale=0.18,angle=270]{nu1.3.eps}
\end{center}
\caption{The density of $\nu_t$ for $V(x)=\frac{x^4}{4}-x^2$ with
$t$ equal to $0.8$, $1$ and $1.3$, respectively (compare with Figure
\ref{figure-tweedeschaling}).} \label{figure-evenwicht}
\end{figure}

Then $\nu_t = \mu_t$ for $t \geq 1$, but for $t < 1$ there
is a clear distinction between $\nu_t$ and $\mu_t$,
see Figure 2. However we still have the analogue
of (\ref{1-BuRa2}) (as we prove)
\begin{equation} \label{1-BuRa3}
    \left. \frac{d }{dt} \left( t\nu_t\right) \right|_{t = 1} = \omega_{S_V},
    \end{equation}
What's more, we also have (\ref{1-BuRa3}) at the level
of densities, that is, if $\psi_t$ denotes the density
of $\nu_t$, and if $x$ is an interior point of $S_V$
(in particular if $x = x^*$), then
\begin{equation} \label{1-BuRa4}
    \left. \frac{d }{dt} \left( t\psi_t(x)\right) \right|_{t = 1}
    = w_{S_V}(x),
    \end{equation}
    where $w_{S_V}$ is the density of the equilibrium
    measure of $S_V$, which is what we need for the proof of Theorem
\ref{theorem-1-maintheorem}.
Note that we do not have (\ref{1-BuRa4}) for $x = x^*$
if $\psi_t$ is the density of $\mu_t$.

\paragraph{Riemann-Hilbert problem.}
The second main tool for the proof of Theorem \ref{theorem-1-maintheorem}
is the characterization of orthonormal polynomials by means
of a Riemann-Hilbert problem, due to Fokas, Its, and Kitaev
\cite{FIK1}, and the subsequent asymptotic analysis of the
Riemann-Hilbert problem by means of the Deift/Zhou steepest
descent analysis of Riemann-Hilbert problems, introduced
in \cite{DZ1}, and further developed in \cite{DVZ,DKMVZ1,DKMVZ2},
and other more recent papers.
The Deift/Zhou steepest descent analysis of the Riemann-Hilbert problem
consists of a sequence of explicit transformations, which result
in a Riemann-Hilbert problem that is explicitly solvable in terms
of Neumann series. (In fact, in this paper, we only need the first
term of this series.) The critical point $x^*$ needs special attention.

Of particular interest for us is the paper \cite{BDJ} by Baik, Deift,
and Johansson on the length of the longest increasing subsequence of
a random permutation of $\{1,2, \ldots, n\}$. These authors show
that the fluctuations of this random variable are distributed according
to the Tracy-Widom distribution \cite{TW} in the limit as $n \to \infty$.
One of the technical tools in this important paper is the asymptotic analysis
of a Riemann-Hilbert problem on the unit circle which is related to
an equilibrium measure (also on the unit circle)
whose density vanishes at the point $-1$. This  situation is
comparable to ours. The authors of \cite{BDJ},
see also subsequent papers  \cite{BDJ2,BR1,BR2,BDR},
construct a local parametrix near $-1$ with the aid of the $\psi$-functions
associated with the Hastings-McLeod solution of Painlev\'e II.
These $\psi$-functions satisfy a model Riemann-Hilbert problem
and the local parametrix is constructed by
appropriately mapping the model Riemann-Hilbert problem
onto a neighborhood of $-1$ so that it satisfies certain
desired jump properties approximately.

We follow the approach of \cite{BDJ} but we introduce
a modification in the construction of the local parametrix
so that it has the desired jump properties {\em exactly},
in contrast to \cite{BDJ} where the desired jump properties
only hold {\em approximately}. The fact that we have the
exact jump properties simplifies the arguments considerably
and we feel that this is also a main contribution
of the present paper.

\paragraph{Outline of the rest of the paper.}

In Section 3 we collect the necessary facts about equilibrium measures.
In particular we study the modified equilibrium problem with external
field in some detail. In Section 4 we discuss the Riemann-Hilbert
problem satisfied by the $\psi$-functions. Here we follow \cite{FN}.
Then in Section 5 we state the Riemann-Hilbert problem
for orthogonal polynomials, discuss the relation with the
correlation kernel $K_{n,N}$, and perform the transformations
in the steepest descent analysis.
Finally in Section 6 we give the proof of
Theorem \ref{theorem-1-maintheorem}.

\section{Equilibrium Measures}

As explained in the previous section, we consider a modification
of the equilibrium problem where we drop the non-negativity condition
in a small neighborhood of $x^*$. We take $\delta_0 > 0$
sufficiently small so that
\begin{equation} \label{delta0}
    \psi_V(x) > 0 \qquad \mbox{ for all }
    x \in [x^*-\delta_0, x^*+\delta_0] \setminus \{x^*\}.
\end{equation}
and we use $\nu_t$ to denote the signed measure that
minimizes (\ref{mod-energy1}) under the conditions
(\ref{mod-energy2}). We define
\begin{equation} \label{def-St}
    S_t = \supp(\nu_t).
\end{equation}
The existence and uniqueness of $\nu_t$ follows
as in \cite{SaTo}.

Let
\[ U^{\nu}(x) = \int \log \frac{1}{|x-y|} d \nu(y) \]
be the logarithmic potential of $\nu$.
Then standard arguments of potential theory
\cite{Deift,SaTo} show that $\nu_t$ is the unique signed
measure satisfying (\ref{mod-energy1}) with the
property that
\begin{align} \label{variational-nut1}
   2U^{\nu_t}(x) + \frac{1}{t} V(x) & =  \ell_t,
        \qquad x \in \supp(\nu_t) \cup [x^* - \delta_0, x^* + \delta_0], \\
        \label{variational-nut2}
   2U^{\nu_t}(x) + \frac{1}{t} V(x) & \geq  \ell_t,
    \qquad x \in \mathbb R.
\end{align}
for some constant $\ell_t$.

As before we use $\mu_t$ to denote the equilibrium measure
in the external field $V_t$. This is a probability measure
that satisfies for some constant $\tilde{\ell}_t$,
\begin{align} \label{variational-mut1}
   2U^{\mu_t}(x) + \frac{1}{t} V(x) & =  \tilde{\ell}_t,
        \qquad x \in \supp(\mu_t), \\
        \label{variational-mut2}
   2U^{\mu_t}(x) + \frac{1}{t} V(x) & \geq  \tilde{\ell}_t,
    \qquad x \in \mathbb R.
\end{align}
It is known that $t\mu_t$ and $\supp(\mu_t)$ are increasing with
$t>0$, see \cite{BR,SaTo}. For $t \geq 1$, we have $\supp(\mu_V)
\subset \supp(\mu_t)$ which implies that $\nu_t = \mu_t$ in view
of (\ref{delta0}) and the variational conditions
(\ref{variational-nut1})--(\ref{variational-mut2}). For $t < 1$,
we have that $x^*$ is outside the support of $\mu_t$, and in fact
the strict inequality in (\ref{variational-mut2}) holds for $x =
x^*$ if $t < 1$.

In general there is the following inequality between $\mu_t$ and $\nu_t$.
\begin{lemma}
For every $t > 0$ we have
\begin{equation} \label{mut-leq-nut}
    \mu_t  \leq  \nu_t^+.
\end{equation}
\end{lemma}
\begin{proof}
Let $\lambda = \nu_t - \mu_t$. From (\ref{variational-nut1}),
(\ref{variational-nut2}), (\ref{variational-mut1}), and (\ref{variational-mut2})
it follows that
\begin{align} \label{Ulambda6}
2 U^{\lambda}(x) & \leq
 \ell_{t}- \tilde{\ell}_{t},\qquad \textrm{ for } x\in S_{t},
\\ \label{Ulambda7}
2 U^{\lambda}(x) & \geq  \ell_{t} - \tilde{\ell}_{t},
\qquad \textrm{ for } x\in \supp(\mu_t).
\end{align}
The potential $U^{\lambda}$ is subharmonic on
$\mathbb C \setminus \supp(\lambda^+)$ and since
$\int d\lambda = 0$, it is
subharmonic at infinity as well. By the  maximum
principle for subharmonic functions \cite{Ransford,SaTo},
the maximum of
$U^{\lambda}$ is attained in $\supp(\lambda^+)$ only.
Since $\supp(\lambda^+) \subset S_t$ we  then have by (\ref{Ulambda6})
that equality  in (\ref{Ulambda7}) holds for every
$x \in \supp(\mu_t)$, and so $\supp(\mu_t) \subset \supp(\lambda^+)$.
This implies (\ref{mut-leq-nut}).
\end{proof}

It follows from (\ref{mut-leq-nut}) that $\supp(\nu_t^-) \cap
\supp(\mu_t)$ is empty. Since for $t$ slightly less than $1$
a gap opens in $\supp(\mu_t)$, which depends continuously on $t$,
see \cite{KM},
it follows that for any given $\delta \in (0, \delta_0)$,
there is $t_0 < 1$ such that
\[ \supp(\nu_t^-) \subset [x^*-\delta, x^*+\delta],
    \qquad \mbox{ for } t > t_0. \]
This shows that for $t < 1$ sufficiently close to $1$,
the definition of $\nu_t$ is independent of the choice of $\delta_0$.

A very useful fact is that for real analytic $V$, say $V$ is
analytic in a neighborhood $\mathcal{V}$ of the real line, the
measures $\mu_t$ have densities $\tilde{\psi}_t$ which can be
expressed in terms of the negative part of an analytic function in
$\mathcal{V}$. Indeed, if
\begin{equation} \label{tilde-qt}
    \tilde{q}_t(z) = \left(\frac{V'(z)}{2t}\right)^2
    - \frac{1}{t} \int \frac{V'(z)-V'(y)}{z-y} d\mu_t(y),
\end{equation}
    then it was shown in \cite{DKM} that
\begin{equation} \label{tilde-psit}
\tilde{\psi}_t(x) = \frac{1}{\pi} \sqrt{\tilde{q}^-_t(x)},
\end{equation}
where $\tilde{q}_t^- = \max(0,-\tilde{q}_t)$ denotes the
negative part of $\tilde{q}_t$.
A consequence of (\ref{tilde-psit}) is that
$\supp(\mu_t)$ is the closure of the set where
$\tilde{q}_t$ is negative.

The arguments of \cite{DKM} can be readily extended to
the signed measures $\nu_t$, provided that
$\supp(\nu_t^-) \subset [x^*-\delta,x^*+\delta]$ for
some $\delta < \delta_0$. We define
\begin{equation} \label{qt}
    q_t(z) = \left(\frac{V'(z)}{2t}\right)^2
    - \frac{1}{t} \int \frac{V'(z)-V'(y)}{z-y} d\nu_t(y),
    \qquad z \in \mathcal V.
\end{equation}
and then the following holds.
\begin{proposition} \label{prop2-1}
There exists $t_0 \in (0,1)$ such that
for every $t > t_0$, the signed measure $\nu_t$ has
a density $\psi_t$ and
\begin{equation} \label{qt1}
    q_t(x) = - \left[\pi \psi_t(x) \right]^2, \qquad \mbox{ for } x \in S_t.
\end{equation}
In addition we have
\begin{equation} \label{qt2}
    q_t(x) = \left[ \int_{S_t} \frac{\psi_t(y)}{y-x} dy + \frac{V'(x)}{2} \right]^2
    \qquad \mbox{ for } x \in \mathbb R \setminus S_t.
\end{equation}
\end{proposition}
\begin{proof}
This follows as in \cite[Proposition 2.51]{DKM}.
\end{proof}

Obviously, for $t \geq 1$ we have $q_t = \tilde{q}_t$.
Since $\psi_1 = \psi_V$ has a double zero at $x^*$, we see
from (\ref{qt1}) that $q_1$ has a zero at $x^*$ of order four.
For $t$ slightly bigger than $1$, this fourth order zero splits into two double zeros
in the complex plane away from the real axis.

For $t$ slightly less than $1$, there is a difference in
the behavior of the zeros of $q_t$ and $\tilde{q}_t$.
Indeed, $\tilde{q}_t$ has two simple real zeros near $x^*$
which are endpoints of the support of $\mu_t$, and
in addition there is a double real zero in between them.
On the other hand, we have that $q_t$ has two double real zeros
near $x^*$, which are endpoints of the support of $\nu_t^-$.

The fact that $-q_t$ has only double zeros near $x^*$
allows us to take an analytic square root, and in
view of (\ref{qt1}) we choose it so that for $z \in S_t$,
\begin{equation} \label{qt3}
    \psi_t(z) = \frac{1}{\pi} (-q_t(z))^{1/2},
\end{equation}
where the sign of the square root at $x = x^*$
is taken negative if $t < 1$ and positive if $t > 1$.
The right-hand side of (\ref{qt3}) has an analytic extension
to a neighborhood of $x^*$, which is independent of the value
of $t > t_0$. Thus $\psi_t$ has an analytic extension to
a fixed neighborhood of $x^*$, which will also
be denoted by $\psi_t$.

Recall from the discussion before the statement of
Theorem 2.1, that the density $w_{S_V}$ of the
equilibrium measure of $S_V$ also has an analytic
extension to a neighborhood of $x^*$, which
we also denote by $w_{S_V}$.
The remaining part of this section is devoted
to the proof of the following proposition.

\begin{proposition} \label{prop2-2}
We have
\begin{equation} \label{diff-tpsi}
    \lim_{t \to 1} \frac{t\psi_t(z) - \psi_1(z)}{t-1} =
    w_{S_V}(z)
    \end{equation}
uniformly for $z$ in a neighborhood of $x^*$.
\end{proposition}

We start with a lemma which contains a weaker form
of (\ref{diff-tpsi}).

\begin{lemma}
We have
    \begin{equation} \label{diff-tnu}
        \left. \frac{d }{dt} (t\nu_t)  \right|_{t=1} = \omega_{S_V}
    \end{equation}
       where $\omega_{S_V}$ is the equilibrium measure of $S_V$.
\end{lemma}

\begin{proof}
Buyarov and Rakhmanov \cite{BR} proved (for a very general class of $V$)
that
\begin{equation} \label{BuRa1}
    \lim_{t \to 1-} \frac{t\mu_t - \mu_V}{t-1} = \omega_{S_V}
\end{equation}
and
\begin{equation} \label{BuRa2}
    \lim_{t \to 1+} \frac{t\mu_t - \mu_V}{t-1} = \omega_{S^*_V}
\end{equation}
where $\omega_{S^*_V}$ is the equilibrium measure of the
set
\[ S^*_V = \{x \in \mathbb R \mid 2 U^{\mu_V}(x) + V(x) = \tilde{\ell}_1 \}, \]
see (\ref{variational-mut1}) with $t=1$. For real analytic $V$, the sets $S_V$
and $S^*_V$ differ by an at most finite number of points, so that
$\omega_{S_V} = \omega_{S^*_V}$. Thus by (\ref{BuRa1}) and (\ref{BuRa2})
we have
\begin{equation} \label{BuRa3}
    \lim_{t \to 1} \frac{t\mu_t - \mu_V}{t-1} = \omega_{S_V}.
\end{equation}

Now write
\[ \frac{t \nu_t - \nu_1}{t-1} =
    \frac{t\mu_t - \mu_V}{t-1} +  \frac{t}{t-1} (\nu_t - \mu_t). \]
In view of (\ref{BuRa3}) it suffices to prove that
\begin{equation} \label{nut-mut}
    \| \nu_t - \mu_t \| = o(t-1) \qquad \mbox{ as } t \to 1,
\end{equation}
where the norm denotes the total variation of a signed measure.
We may assume that $t < 1$.
Because of (\ref{mut-leq-nut}) we have
\begin{equation} \label{nut-mut2}
    \| \nu_t - \mu_t \| = \| \nu_t^+ - \mu_t - \nu_t^- \|
    = \int d(\nu_t^+ - \mu_t) + \int d\nu_t^-
    = 2 \int d \nu_t^-.
    \end{equation}
For $t < 1$, the support of $\nu_t^-$ is contained in
an interval $[x^*-\delta(t), x^*+\delta(t)]$, with
\begin{equation} \label{deltat1}
    \delta(t) = O\left((1-t)^{1/2}\right)
    \qquad \mbox{ as } t \to 1-,
\end{equation}
see \cite[Lemma 8.1(iii)]{KM}.
From (\ref{qt3}) and the fact that $q_t$ has
two double zeros in $[x^*-\delta(t), x^*+\delta(t)]$
which are the end-points of the support of $\nu_t^-$,
we then easily get that
\begin{equation} \label{deltat2}
    \frac{d}{dx} \nu_t^-(x) = O(\delta(t)^2) = O(1-t)
    \qquad \mbox{ as } t \to 1-.
\end{equation}
Combining (\ref{deltat1}) and (\ref{deltat2}), we find
$\int d\nu_t^- = O\left((1-t)^{3/2}\right)$ as $t \to 1-$,
which by (\ref{nut-mut2}) implies (\ref{nut-mut}).
This completes the proof of the lemma.
\end{proof}

We now give a characterization of $\psi_V$, which will be of use
in the proof of Proposition \ref{prop2-2}.

\begin{lemma}\label{lemmaclaim}
For $x \in S_V$ we have
\begin{equation} \label{eq:psiV-lemma}
    \psi_V(x) = \frac{1}{2\pi^2} \frac{1}{w_{S_V}(x)}
    \int \frac{V'(x) - V'(y)}{x-y} d\omega_{S_V}(y),
         \end{equation}
where $w_{S_V}$ is the density of $\omega_{S_V}$.
\end{lemma}
\begin{proof}
By the Sokhotski-Plemelj formulas, see e.g.\ \cite{Gakhov},
we have that
\[ F(z)=\int_{S_V}\frac{\psi_V(s)}{z-s}ds,
    \qquad \mbox{ for } z \in \mathbb C \setminus S_V, \]
satisfies
\begin{eqnarray} \label{SP1}
F_+(x)+F_-(x) & = &V'(x),  \qquad \qquad x\in S_V, \\
\label{SP2}
F_+(x)-F_-(x) & = & -2\pi i\psi_V(x), \qquad x\in S_V.
\end{eqnarray}
Let $S_V=\bigcup_{j=1}^n[a_j,b_j]$ and set
\[R(z)=\left(\prod_{j=1}^n(z-b_j)(z-a_j)\right)^{1/2},
    \qquad z \in \mathbb C \setminus S_V, \]
where the square root  is positive for $z > b_n$.
Using (\ref{SP1}) and the fact that $R_+(x)=-R_-(x)$ as $x\in
S_V$, we see that
\begin{equation}\frac{p(x)F_+(x)}{R_+(x)}-\frac{p(x)F_-(x)}{R_-(x)}=
\frac{p(x)V'(x)}{R_+(x)}\label{sprongpFR}\end{equation} for any
polynomial $p$. Suppose $p$ has degree at most $n-1$.
Then (\ref{sprongpFR}) and the fact that $\frac{p(z)F(z)}{R(z)}\to
0$ as $z\to \infty$ imply  that
\begin{equation*}
    \frac{p(z)F(z)}{R(z)}=\frac{1}{2\pi i}
    \int_{S_V}\frac{p(s)V'(s)}{R_+(s)}\frac{1}{s-z}ds,
    \qquad z \in \mathbb C \setminus S_V
\end{equation*}
which we can rewrite for $z \in \mathcal V$ as
\begin{align} \nonumber
    \frac{p(z) F(z)}{R(z)} & = \frac{1}{2\pi i}
    \int_{S_V}\frac{V'(s)-V'(z)}{s-z}\frac{p(s)}{R_+(s)}ds+
    \frac{V'(z)}{2\pi i}
    \int_{S_V} \frac{p(s)}{R_+(s)} \frac{1}{s-z} ds \\
    & = \label{eq:Fz-lemma}
    \frac{1}{2\pi i}
    \int_{S_V}\frac{V'(z)-V'(s)}{z-s}\frac{p(s)}{R_+(s)}ds
    +
    \frac{V'(z)}{2} \frac{p(z)}{R(z)}
    \end{align}
by contour integration. Now we use
(\ref{eq:Fz-lemma}) and (\ref{SP2}) to obtain
\begin{equation*}
    \psi_V(x) = \frac{F_+(x)-F_-(x)}{-2\pi i}
    =\frac{1}{2\pi^2}\frac{R_+(x)}{p(x)}
    \int_{S_V}\frac{V'(x)-V'(y)}{x-y}\frac{p(y)}{R_+(y)}dy,
\end{equation*}
and this holds for any polynomial $p$ of degree at most $n-1$.
Since we know by (\ref{wSdensity}) that
\[ w_{S_V}(x)=\frac{i}{\pi}\frac{p(x)}{R_+(x)}, \qquad x\in S_V, \]
for some monic polynomial $p$ of degree $n-1$,
we get (\ref{eq:psiV-lemma}).
\end{proof}

Now we can give the proof of Proposition \ref{prop2-2}.

\begin{varproof} {\bf of Proposition \ref{prop2-2}.}
By (\ref{qt}) we have for $z \in \mathcal V$,
\[ t^2 q_t(z) = \left(\frac{V'(z)}{2} \right)^2
    - \int \frac{V'(z)-V'(y)}{z-y} d (t\nu_t)(y) \]
so that in view of (\ref{diff-tnu}) we find,
\[ \left. \frac{d}{dt}
    (t^2 q_t(z)) \right|_{t=1} =
    - \int \frac{V'(z) - V'(y)}{z-y} d \omega_{S_V}(y). \]
Thus by (\ref{qt3}) for $z \neq x^*$ in a neighborhood
of $x^*$,
\begin{align} \nonumber
    \left. \frac{d}{d t}  (t\psi_t(z)) \right|_{t=1}
    & = \left. \frac{d}{d t} \nonumber
         \frac{1}{\pi} (-t^2 q_t(z))^{1/2} \right|_{t=1} \\
    & = \frac{1}{2\pi} (-q_V(z))^{-1/2} \nonumber
     \int \frac{V'(z) - V'(y)}{z-y} d \omega_{S_V}(y) \\
    & = \frac{1}{2\pi^2} \frac{1}{\psi_V(z)}
    \int \frac{V'(z)-V'(y)}{z-y} d\omega_{S_V}(y).
        \label{diff-tpsi2}
    \end{align}
By (\ref{eq:psiV-lemma}), the right-hand side of
(\ref{diff-tpsi2}) is $\omega_{S_V}(z)$ in case
$z \neq x^*$ is real, and by analytic continuation it continues
to be $\omega_{S_V}(z)$ in a punctured neighborhood of $x^*$.
Thus (\ref{diff-tpsi}) holds for $z \neq x^*$ in a neighborhood
of $x^*$, and then it easily follows for $z = x^*$ as well.
\end{varproof}

\section{Riemann-Hilbert Problem Associated with
the Painlev\'e II Equation}
In this section we recall the RH
problem associated with the second Painlev\'e equation, see
\cite{FN,IN,DZ2,BDJ,IK}. We transform this RH problem into a RH
problem that will be used in the next section to construct a
parametrix around $x^*$.
\subsection{$\psi$-functions Associated with Painlev\'e II}
Consider the differential equation of (\ref{1-diiffvglpsi1}),
\begin{equation}
\frac{d}{d\zeta} \Psi(\zeta)=
\begin{pmatrix}
 -4i\zeta^2-i(s+2q^2)  & 4\zeta q+2ir \\
4\zeta q -2ir   & 4i\zeta^2+i(s+2q^2)
\end{pmatrix} \Psi(\zeta),\label{4-diffvglpsi}
\end{equation} where $\Psi$
is a $2\times 2$ complex matrix-valued function and
$s$, $q$ and $r$ are considered as parameters.
All solutions of (\ref{4-diffvglpsi}) are entire functions
of $\zeta$.

For $j=1, \ldots, 6$, let $S_j$ be the sector
\begin{equation} \label{sector}
S_j=\{\zeta\in\mathbb C \mid
    \frac{2j-3}{6} \pi < \arg \zeta <
    \frac{2j-1}{6} \pi \}.
\end{equation}
There exists a unique solution $\Psi_j$ of equation (\ref{4-diffvglpsi}) so that
\begin{equation} \label{Psij}
 \Psi_j(\zeta) e^{i(\frac{4}{3}\zeta^3+s\zeta)\sigma_3} = I+O(\zeta^{-1})
\end{equation}
as $\zeta\to \infty$ in the sector $S_j$. Here we use
$\sigma_3 = \begin{pmatrix} 1 & 0 \\ 0 & -1 \end{pmatrix}$
to denote the third Pauli matrix.
There exist complex values $a_j$, $j=1, \ldots, 6$
(called Stokes multipliers) so that
\begin{equation} \label{Psijumps}
    \Psi_{j+1}(\zeta)=\Psi_j(\zeta)A_j, \quad
    \mbox{for } j =1, \ldots, 5, \qquad
    \Psi_1(\zeta) = \Psi_6(\zeta) A_6.
\end{equation}
with
\begin{equation} \label{Ajodd}
    A_j=    \begin{pmatrix}
  1 & 0 \\   a_j & 1  \end{pmatrix}
 \quad \textrm{ if $j$ is odd,}
\end{equation}
and
\begin{equation} \label{Ajeven}
    A_j= \begin{pmatrix}
  1 & a_j \\   0 & 1 \\ \end{pmatrix}
  \quad \textrm{ if $j$ is even.}
\end{equation}
Furthermore, we have
\begin{equation}
 a_{j+3} = a_j  \qquad \mbox{and} \qquad
 a_1 a_2 a_3 + a_1 + a_2 + a_3 = 0.
\end{equation}
Now define the rays
\begin{equation} \label{Gammaj}
\Gamma_j = \{ \zeta \mid \arg \zeta = \frac{2j-1}{6} \pi \}
\quad \textrm{ for }j=1, \ldots, 6,
\end{equation}
oriented away from the origin, and the matrix-valued
function
\begin{equation} \label{def-Psi}
\Psi(\zeta) := \Psi_j(\zeta), \qquad \mbox{ for } \zeta \in S_j.
\end{equation}
Then $\Psi$ satisfies the following RH problem
\begin{itemize}
\item[(a)] $\Psi$ is analytic on
$\mathbb C \setminus \bigcup_{j=1}^6\Gamma_j$.
\item[(b)] $\Psi_+ = \Psi_- A_j$ on $\Gamma_j$,
\item[(c)] $\Psi(\zeta) e^{i(\frac{4}{3}\zeta^3+s\zeta)\sigma_3}=I+O(\zeta^{-1})$
as $\zeta\to \infty$,
\item[(d)] $\Psi$ is bounded near $0$.
\end{itemize}

\begin{figure}[t] \label{fig:jumpsPsialgemeen}
\begin{center}
\setlength{\unitlength}{0.8truemm}
\begin{picture}(100,100)(0,0)
\put(77,57){{\small $\begin{pmatrix}1&0\\
a_1&1\end{pmatrix}$}}
\put(51,80){{\small $\begin{pmatrix}1&a_2\\
0&1\end{pmatrix}$}}
\put(14,72){{\small $\begin{pmatrix}1&0\\
a_3&1\end{pmatrix}$}}
\put(7,41){{\small $\begin{pmatrix}1&a_1\\
0&1\end{pmatrix}$}}
\put(33,18){{\small $\begin{pmatrix}1&0\\
a_2&1\end{pmatrix}$}}
\put(68,26){{\small $\begin{pmatrix}1&a_3\\
0&1\end{pmatrix}$}}
\put(50,50){\vector(0,1){23}} \put(50,50){\vector(0,-1){23}}
\put(50,73){\line(0,1){27}} \put(50,27){\line(0,-1){27}}
\put(50,50){\vector(2,1){23}} \put(50,50){\vector(2,-1){23}}
\put(73,61.5){\line(2,1){27}} \put(73,38.5){\line(2,-1){27}}
\put(50,50){\vector(-2,1){23}} \put(50,50){\vector(-2,-1){23}}
\put(27,61.5){\line(-2,1){27}} \put(27,38.5){\line(-2,-1){27}}

\end{picture}
\end{center}
\caption{Jumps for $\Psi(\zeta;s)$.}
\end{figure}
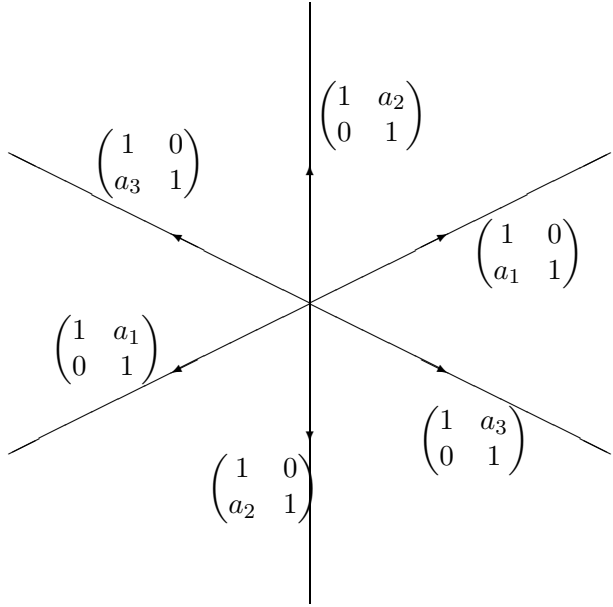

The Stokes multipliers $a_j$ depend on $s$, $q$ and $r$. An
isomonodromy deformation is a variation of these parameters such
that the Stokes multipliers remain constant. Flaschka and Newell
\cite{FN} showed that the isomonodromy deformations are given by
the Painlev\'e II equation $q''(s) = sq(s) + 2q^3(s)$ and $r(s) =
q'(s)$.

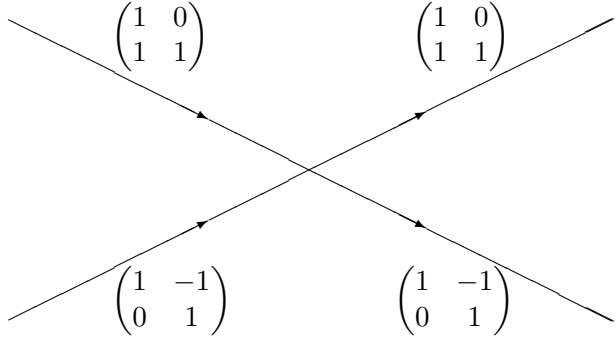
\begin{figure}[t]
\begin{center}
\setlength{\unitlength}{0.8truemm}
\begin{picture}(100,50)(0,0)
\put(17,2){{\small $\begin{pmatrix}1&-1\\
0&1\end{pmatrix}$}}
\put(17,46){{\small $\begin{pmatrix}1&0\\
1&1\end{pmatrix}$}}
\put(64,2){{\small $\begin{pmatrix}1&-1\\
0&1\end{pmatrix}$}}
\put(67,46){{\small $\begin{pmatrix}1&0\\
1&1\end{pmatrix}$}}
\put(0,0){\vector(2,1){33.1}} \put(33.33,16.665){\vector(2,1){36}}
\put(69.5,34.72){\line(2,1){31}}
\put(0,50){\vector(2,-1){33.1}}
\put(33.33,33.33){\vector(2,-1){36}}
\put(69.5,15.275){\line(2,-1){31}}
\end{picture}
\end{center}
\caption{Jumps for $\Psi(\zeta;s)$.}\label{fig:jumpsPsi}
\end{figure}

Any solution of the Painlev\'e II equation is a meromorphic
function with an infinite number of poles. We write
$\Psi(\zeta;s)$ for the $\Psi$ function (\ref{def-Psi}) with
parameters $s$, $q = q(s)$, $r= r(s)$, where $q$ is the
Hastings-McLeod solution and $r(s) = q'(s)$. Let $\cal P$ be the
set of poles of $q$. Then $\Psi(\zeta;s)$ is defined and analytic
for $\zeta \in \mathbb C \setminus \bigcup_{j=1}^6 \Gamma_j$ and
for $s \in \mathbb C \setminus \mathcal P$. It is known that there
are no poles on the real line \cite{HM}. The Stokes multipliers
corresponding to the Hastings-McLeod solution are
\[ a_1 = 1, \quad a_2 = 0, \quad a_3 = -1. \]
Thus $\Psi(\zeta;s)$ is analytic across the
imaginary axis. We reverse the orientation of $\Gamma_3$ and
$\Gamma_4$, and we define
\[ \Sigma_1 = \Gamma_1 \cup \Gamma_3,
    \quad \mbox{ and } \quad \Sigma_2 = \Gamma_4 \cup \Gamma_6.
    \]
Then $\Psi(\zeta;s)$ solves the following RH problem,
see also Figure \ref{fig:jumpsPsi}.
\begin{itemize}
\item[(a)] $\Psi$ is analytic on
$\mathbb C \setminus (\Sigma_1 \cup \Sigma_2)$,
\item[(b1)] $\Psi_+ = \Psi_-
    \begin{pmatrix} 1 & 0 \\ 1 & 1 \end{pmatrix}$
     on $\Sigma_1$,\\
  \vspace{-9.5mm}\begin{equation}\label{RHPPsib1}\end{equation}
\item[(b2)] $\Psi_+ = \Psi_- \begin{pmatrix} 1 & -1 \\ 0 & 1
    \end{pmatrix}$ on $\Sigma_2$,
\item[(c)] $\Psi(\zeta;s)
e^{i(\frac{4}{3}\zeta^3+s\zeta)\sigma_3}=I+O(\zeta^{-1})$ as
$\zeta\to \infty$, \item[(d)] $\Psi$ is bounded near $0$.
\end{itemize}
The RH problem has a solution if and only if
$s \in \mathbb C \setminus \mathcal P$. The properties
(c) and (d) are valid uniformly for $s$ in compact
subsets of $\mathbb C \setminus \mathcal P$.

\subsection{Transformation of the RH problem}
We modify the RH problem for $\Psi$ so that it resembles
the RH problem that we will need locally near $x^*$.

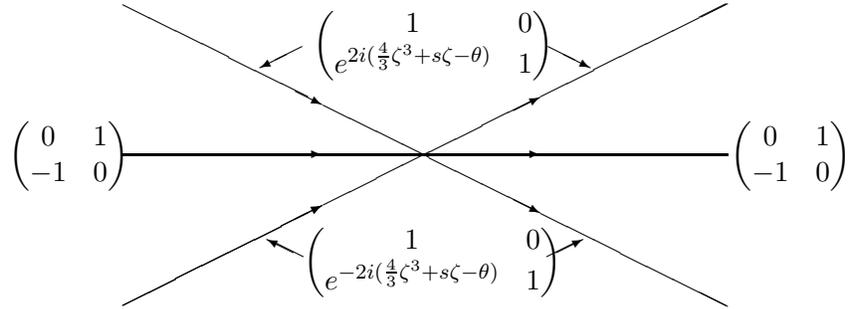
\begin{figure}[t]
\begin{center}
\setlength{\unitlength}{0.8truemm}
\begin{picture}(140,50)(-20,0)
\put(30,6){{\small $\begin{pmatrix}1&0\\
e^{-2i(\frac{4}{3}\zeta^3+s\zeta-\theta)}&1\end{pmatrix}$}}
\put(31.6,42){{\small $\begin{pmatrix}1&0\\
e^{2i(\frac{4}{3}\zeta^3+s\zeta-\theta)}&1\end{pmatrix}$}}
\put(-19,23.6){{\small $\begin{pmatrix}0&1\\
-1&0\end{pmatrix}$}}
\put(101,23.6){{\small $\begin{pmatrix}0&1\\
-1&0\end{pmatrix}$}} \put(0,0){\vector(2,1){33.1}}
\put(30,8){\vector(-2,1){6}} \put(70.5,8){\vector(2,1){6}}
\put(29.8,43){\vector(-2,-1){7}}\put(70.7,43){\vector(2,-1){7}}
\put(0,0){\vector(2,1){33.1}} \put(33.33,16.665){\vector(2,1){36}}
\put(69.5,34.72){\line(2,1){31}}
\put(0,50){\vector(2,-1){33.1}}
\put(33.33,33.33){\vector(2,-1){36}}
\put(69.5,15.275){\line(2,-1){31}}
\put(0,25){\vector(1,0){33.1}} \put(33.33,25){\vector(1,0){36}}
\put(69.5,25){\line(1,0){31}}
\end{picture}
\end{center}
\caption{Jumps for $M(\zeta;s,\theta)$.}\label{fig:jumpsM}
\end{figure}

We introduce an additional parameter $\theta \in \mathbb R$
and define
\begin{equation} \label{def-M}
M(\zeta;s,\theta) =
    \begin{cases}
    \begin{array}{ll}
    e^{i \theta \sigma_3}\Psi(\zeta;s)e^{i(\frac{4}{3}\zeta^3+s\zeta)\sigma_3}
    e^{-i \theta \sigma_3} &
    \mbox{for } \Im \zeta > 0, \\
    e^{i\theta \sigma_3}\Psi(\zeta;s)e^{i(\frac{4}{3}\zeta^3+s\zeta)\sigma_3}
    e^{-i \theta \sigma_3}
    \begin{pmatrix} 0 & -1 \\ 1 & 0 \end{pmatrix} &
    \mbox{for } \Im \zeta < 0.
    \end{array} \end{cases}
\end{equation}
For any given $s \in \mathbb C \setminus \mathcal P$ and $\theta
\in \mathbb R$, we then have that $M$ is defined for $\zeta \in
\mathbb C \setminus (\mathbb R \cup \Sigma_1 \cup \Sigma_2)$, see
Figure \ref{fig:jumpsM}, and satisfies the following RH problem
\begin{itemize}
\item[(a)] $M(\zeta;s, \theta)$ is analytic for
$\zeta \in \mathbb C \setminus(\mathbb R \cup \Sigma_1 \cup \Sigma_2)$
and $s \in \mathbb C \setminus \mathcal P$.
\item[(b1)] $M_+(\zeta;s,\theta)= M_-(\zeta;s,\theta)
\begin{pmatrix}
  1 & 0 \\
  e^{2i(\frac{4}{3}\zeta^3+s\zeta- \theta)} & 1
\end{pmatrix}$ for $\zeta \in \Sigma_1$,
\item[(b2)] $M_+(\zeta;s,\theta) = M_-(\zeta;s,\theta)
\begin{pmatrix}
  0 & 1 \\
  -1 & 0
\end{pmatrix}$ for $\zeta\in\mathbb R$,
\item[(b3)] $M_+(\zeta;s,\theta) = M_-(\zeta;s,\theta)
\begin{pmatrix}
  1 & 0 \\
  e^{-2i(\frac{4}{3}\zeta^3+s\zeta-\theta)} & 1
\end{pmatrix}$ for $\zeta\in\Sigma_2$.
\item[(c1)]
$M(\zeta;s,\theta)=I+O(\zeta^{-1})$  as $\zeta \to
\infty$ in the upper half plane,\\
  \vspace{-9.5mm}\begin{equation}\label{RHPMc1}\end{equation}
\item[(c2)]
$M(\zeta;s,\theta)=(I+O(\zeta^{-1}))\begin{pmatrix}
  0 & 1 \\ -1 & 0 \end{pmatrix} $
  as $\zeta \to \infty$ in the lower
half plane,\\
  \begin{equation}\label{RHPMc2}\end{equation}
\item[(d)]
$M(\zeta;s,\theta)$ is bounded for $\zeta$ near $0$.
\end{itemize}
The properties (c1), (c2), and (d) hold uniformly
for $s$ in compact subsets of $\mathbb C
\setminus \mathcal P$ and for $\theta \in \mathbb R$.

\section{Riemann-Hilbert Analysis}
The proof of Theorem \ref{theorem-1-maintheorem}
is based on the steepest descent analysis of the
Riemann-Hilbert problem for orthogonal polynomials.

Since the main point of the present discussion is the
treatment of the critical point $x^*$, we will
restrict ourselves to the one-interval case.
We also assume that there are no other singular points
besides $x^*$. Thus $S_V = [a,b]$ and $\psi_V$
vanishes like a square root at $a$ and $b$, and
quadratically at $x^*$ but at no other
points of $S_V$. In addition, we assume that the
inequality in the variational condition (\ref{variational-mut2})
is strict for $x \in \mathbb R \setminus [a,b]$.
We then have that
\begin{equation} \label{eq:St-interval}
    S_t = \supp(\nu_t) = [a_t, b_t]
\end{equation}
for certain $a_t < b_t$ for every $t$ close to $t = 1$.
Note that $a_t$ is  an increasing and $b_t$
a decreasing function of $t$.

We will comment below on the modifications that have to
be made in the multi-interval case, see Remark
\ref{remark-5-vereenvoudiging}.

\subsection{RH problem for Orthogonal Polynomials}
For each $n$ and $N$, we consider a Riemann-Hilbert problem,
introduced by Fokas, Its and Kitaev \cite{FIK1,FIK2}. We
will look for a $2\times 2$ matrix-valued function $Y = Y_{n,N}$
(we drop the subscripts for simplicity) that satisfies the following
conditions:
\begin{itemize}
\item[(a)] $Y$ is analytic in $\mathbb C \setminus \mathbb R$,
\item[(b)] $Y_+(x)=Y_-(x) \begin{pmatrix}
  1 & e^{-NV(x)} \\
  0 & 1 \end{pmatrix}$ for $x \in\mathbb R$,\\
  \vspace{-9.5mm}\begin{equation}\label{RHPYb}\end{equation}
\item[(c)] $Y(z)=\left(I+O(z^{-1})\right)
\begin{pmatrix}
  z^n & 0 \\
  0 & z^{-n}
\end{pmatrix}$ \quad  as $z\to \infty$.
\end{itemize}
Here $Y_+(x)$ (resp. $Y_-(x)$) denotes the limit as we approach $x
\in \mathbb R$ from the upper (resp.\ lower) half-plane. The RH
problem possesses a unique solution which is given by
\begin{equation}Y(z)= \begin{pmatrix}
  \kappa_{n,N}^{-1} p_{n,N}(z) & \ds
    \kappa_{n,N}^{-1} \frac{1}{2\pi i}\int_{\mathbb R}
    \frac{p_{n,N}(s)e^{- N V(s)}}{s-z}ds \\[10pt]
  -2\pi i \kappa_{n-1,N} p_{n-1,N}(z)
  & \ds -\kappa_{n-1,N} \int_{\mathbb R}\frac{p_{n-1,N}(s) e^{- N V(s)}}{s-z}ds
\end{pmatrix} \label{3-FIK}
\end{equation}
where $p_{n,N}(x) = \kappa_{n,N} x^n + \cdots $ denotes the orthonormal polynomial
as before, see also \cite{Deift,Kuij}.

The correlation kernel (\ref{1-kern}) can be expressed directly in terms
of the solution of the RH problem. Indeed, by the Christoffel-Darboux formula for
orthogonal polynomials
\[ K_{n,N}(x,y) = e^{-\frac{N}{2}V(x)} e^{-\frac{N}{2}V(y)}
    \frac{\kappa_{n-1,N}}{\kappa_{n,N}}
    \frac{p_{n,N}(x) p_{n-1,N}(y) - p_{n-1,N}(x) p_{n,N}(y)}{x-y}
    \]
which involves the orthogonal polynomials of degrees $n$ and $n-1$ only.
Using (\ref{3-FIK}) and the fact that $\det Y(z) = 1$ for
every $z \in \mathbb C \setminus \mathbb R$, we then get
\begin{equation}\label{3-KnNY} K_{n,N}(x,y) =
e^{-\frac{N}{2}V(x)} e^{-\frac{N}{2}V(y)}
\frac{1}{2\pi i(x-y)}
    \begin{pmatrix} 0 & 1 \end{pmatrix}
    Y_{+}^{-1}(y) Y_{+}(x)
    \begin{pmatrix} 1 \\ 0 \end{pmatrix}.
\end{equation}
Applying the Deift-Zhou steepest-descent method to the RH problem
will allow us to find the asymptotics of $K_{n,N}$ as given
in Theorem \ref{theorem-1-maintheorem}.

We note that an expression like (\ref{3-KnNY}) was first given
in the context of random matrices with external source \cite{ABK,BK1,BK2}
where the eigenvalue correlation kernel is given in terms of
the solution of a $3 \times 3$ matrix valued RH problem, and
the expression similar to (\ref{3-KnNY}) was found to be very convenient
for asymptotic analysis. Also in the present $2\times 2$-case we
find it helpful to work with (\ref{3-KnNY}).

\subsection{Normalization of the RH problem}
In the first transformation we normalize the RH problem at
infinity. A standard approach would be to use the equilibrium
measure $\mu_{t}$ in external field $V_t$ where $t = n/N$. If we
would do this for the case where $t<0$, we would have an
equilibrium measure with a gap in the support around $x^*$, and an
annoying consequence is that the equality in the variational
conditions is not valid near $x^*$. For this reason we have
introduced the signed measures $\nu_t$ in Section 2 and we
will use these measures now to normalize the RH problem.

We let $t = n/N$ and assume $t$ is sufficiently close
to $1$ so that (\ref{eq:St-interval}) holds.
As before we use
\begin{equation} \label{def-psit}
  \psi_{t} := \frac{d\nu_t}{dx}
\end{equation}
to denote the density of $\nu_t$. Then we define the $g$-function
\begin{equation} \label{def-gt}
g_t(z)= \int \log(z-s) d\nu_t(s) = \int \log (z-s)\psi_t(s)ds,
\end{equation}
where we take the branch cut of the logarithm along the negative
real axis. Then the following properties of $g_t$ are easy to check.
\begin{itemize}
\item[\rm (i)] $e^{g_t(z)}$ is analytic in $\mathbb C \setminus
[a_t,b_t]$,
\item[\rm (ii)] $e^{g_t(z)}=z +O(\frac{1}{z})$ \quad  as $z \to \infty$.
\item[\rm (iii)]${g_t}_{+}(x)-{g_t}_-(x) = 2 \pi i \int_x^{b_t}
 \psi_t(s)ds$ \quad for $x\in\mathbb R$,
\item[\rm (iv)] ${g_t}_{+}(x)+{g_t}_-(x)- \frac{1}{t} V(x)+ \ell_t
\leq 0 $ for $x\in\mathbb R\setminus [a_t,b_t]$,
\item[\rm (v)]
${g_t}_{+}(x)+{g_t}_-(x)-\frac{1}{t} V(x)+ \ell_t=0$ for
    $x\in [a_t,b_t]$.
\end{itemize}

In terms of the analytic function $q_t$, see (\ref{qt})
and Proposition \ref{prop2-1}, we have
\begin{equation} \label{gt-prop1}
    {g_t}_+(x) - {g_t}_-(x) = -2 \int_{b_t}^x (q_t(s))^{1/2}_+ ds,
    \qquad \mbox{for } x \in [a_t,b_t],
\end{equation}
    and
\begin{equation} \label{gt-prop2}
    {g_t}_{+}(x)+{g_t}_-(x)- \frac{1}{t} V(x)+ \ell_t
    = -2 \int_{b_t}^x (q_t(s))^{1/2} ds \qquad
    \mbox{for } x > b_t,
\end{equation}
\begin{equation} \label{gt-prop3}
    {g_t}_{+}(x)+{g_t}_-(x)- \frac{1}{t} V(x)+ \ell_t
    = -2 \int_{a_t}^x (q_t(s))^{1/2} ds \qquad
    \mbox{for } x < a_t
\end{equation}
where $(q_t(s))^{1/2}$ is analytic for $s \in \mathcal V \setminus [a_t,b_t]$
and that  square root is taken which is positive for large real $s$.
We define
\begin{equation} \label{def-phi}
    \varphi_t(z) = \int_{b_t}^z (q_t(s))^{1/2} ds
    \end{equation}
and
\begin{equation} \label{def-varphi}
    \tilde{\varphi}_t(z) = \int_{a_t}^z (q_t(s))^{1/2} ds
\end{equation}
which are defined and analytic in the neighborhood $\mathcal{V}$
of the real line where $V$ is analytic with
 cuts along $(-\infty, b_t)$ and
 $(a_t, +\infty)$, respectively.
Note that $q_t$ has simple zeros in $a_t$ and $b_t$ and
only double zeros in $\mathcal V \setminus [a_t,b_t]$,
so that $q_t^{1/2}$ is indeed analytic there. It is possible that
$q_t$ has double real zeros, and that $q_t^{1/2}$ has sign changes
in $(-\infty,a_t)$ or $(b_t,\infty)$. However, by (\ref{gt-prop2})
and (\ref{gt-prop3}) we have that $\varphi_t(x) > 0$ for $x > b_t$
and $\tilde{\varphi}_t(x) > 0$ for $x < a_t$,
since we are in a situation with strict inequality in
(\ref{variational-nut2}) for $x > b_t$ and $x < a_t$.

We will now perform the first transformation of the RH problem: we
take $t = n/N$ and define
\begin{equation}\label{3-TY}
T(z)=e^{\frac{n}{2} \ell_t \sigma_3}\, Y(z) e^{-ng_t(z)\sigma_3}
e^{-\frac{n}{2}\ell_t\sigma_3} \quad \textrm{ for } z\in \mathbb C\setminus\mathbb R.
\end{equation}
Using the jump condition (\ref{RHPYb}) of $Y$ and (\ref{3-TY}), we
easily check that $T_+(x) =T_-(x) J_T(x)$
for $x \in \mathbb R$, where
\begin{align} \nonumber
J_T(x) & = e^{\frac{n}{2} \ell_t \sigma_3}
   e^{n {g_t}_-(x)\sigma_3} \begin{pmatrix}
  1 & e^{-NV(x)} \\ 0 & 1  \end{pmatrix}
  e^{-n{g_t}_+(x)\sigma_3} e^{-\frac{n}{2} \ell_t \sigma_3}\\
  & = \label{def-VT}
  \begin{pmatrix}
  e^{-n({g_t}_+(x)-{g_t}_-(x))} & e^{n({g_t}_+(x)+{g_t}_-(x)-\frac{1}{t} V(x)+ \ell_t)} \\
  0 & e^{n({g_t}_+(x)-{g_t}_-(x))}.
\end{pmatrix}
\end{align}
Because of the properties (\ref{gt-prop1})--(\ref{gt-prop3})
and the definitions (\ref{def-phi})--(\ref{def-varphi}),
we see that the jump matrix $J_T$ has the following forms on the respective
intervals $(a_t, b_t)$, $(b_t,\infty)$, and $(-\infty, a_t)$,
\[ \begin{aligned}
J_T & = \begin{pmatrix}
    e^{2n {\varphi_t}_+} & 1 \\
    0 & e^{2n {\varphi_t}_-} \end{pmatrix}
    \quad \textrm{ on } (a_t,b_t) \\
J_T & = \begin{pmatrix}
    1 & e^{-2n \varphi_t} \\
    0 & 1 \end{pmatrix} \quad
    \textrm{ on } (b_t, \infty) \\
J_T & = \begin{pmatrix}
    1 & e^{-2n \tilde{\varphi}_t} \\
    0 & 1 \end{pmatrix} \quad
    \textrm{ on } (-\infty, a_t).
    \end{aligned} \]
Thus $T$ is the unique solution of the RH problem
\begin{itemize}
\item[(a)] $T$ is analytic in $\mathbb C\setminus\mathbb R$,
\item[(b)] $T_+(x)=T_-(x)J_T(x)$\quad  for $x\in \mathbb R$,
\item[(c)] $T(z)=I+O(z^{-1})\quad \textrm{ as }z \to \infty$.
\end{itemize}

\subsection{Opening of the lens}

The jump matrix $J_T$ on the interval $(a_t,b_t)$
has the factorization
\begin{align}
J_T &=  \begin{pmatrix}
  e^{2n {\varphi_t}_+} & 1 \\
  0 & e^{2n{\varphi_t}_-}
\end{pmatrix} \nonumber\\
&=\begin{pmatrix}
  1 & 0 \\
  e^{2n{\varphi_t}_-} & 1
\end{pmatrix}
\begin{pmatrix}
  0 & 1 \\
  -1 & 0
\end{pmatrix}
\begin{pmatrix}
  1 & 0 \\
  e^{2n{\varphi_t}_+} & 1
\end{pmatrix}. \label{3-factorisatie1}
\end{align}

\begin{figure}[h]
\begin{center}
\includegraphics[scale=0.6]{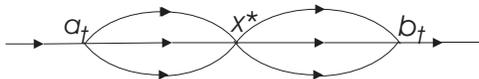}
\end{center}
\caption{The contour after the opening of the lens.}
\label{figure-lens}
\end{figure}

Now we open lenses around the intervals $(a_t,x^*)$ and
$(x^*,b_t)$ as shown in Figure \ref{figure-lens}. Let $C_1$ be the
upper lips of the lenses and $C_2$ the lower lips, with
orientation as in Figure \ref{figure-lens}. We open the
lenses in such a way that they are fully contained in
$\mathcal{V}$, the region of analyticity of $V$ and $q_t$.
In addition, we can take $C_1$ and $C_2$ in such a way that
$\Re \varphi_t < 0$ on $C_1$ and $C_2$, with
the exception of a neighborhood of $x^*$ if $t < 1$.
This effect is due to the fact that $\nu_t$ has a negative
density near $x^*$ if $t < 1$. However, if $t < 1$ increases
to one, the exceptional neighborhood shrinks to a point.
It follows that for any given $\delta > 0$, there
is a constant $\gamma > 0$ such that for $t$ sufficiently
close to $1$ we have
\[ \Re \varphi_t(z) < -\gamma < 0 \]
for all $z \in C_1 \cup C_2$ with $\min(|z-x^*|,|z-a|,|z-b|) > \delta$.

We define
\begin{equation}\label{3-ST}
S=\begin{cases}
    \begin{array}{ll} T &\textrm{ outside the lenses,} \\
    T\begin{pmatrix}
  1 & 0\\
  -e^{2n\varphi_t} & 1
\end{pmatrix} &\textrm{ in upper parts of the lenses,}\\
    T\begin{pmatrix}
    1 & 0 \\
    e^{2n\varphi_t}& 1
\end{pmatrix} &\textrm{ in lower parts of the lenses.}
\end{array}\end{cases}
\end{equation}
The RH problem for $T$ and the factorization
(\ref{3-factorisatie1}) imply that $S$ solves the following RH
problem:
\begin{itemize}
\item[(a)] $S$ is analytic in
$\mathbb C\setminus\left(\mathbb R\cup C_1 \cup C_2\right),$
\item[(b)] $S_+(z)=S_-(z)J_S(z)$ \quad  for $z\in \mathbb R\cup
C_1\cup C_2$,\\
  \vspace{-9.5mm}\begin{equation}\label{RHPSinftyb}\end{equation}
\item[(c)] $S(z)=I+O(z^{-1})$ \quad as $z\to \infty$,
\item[(d)] $S$ remains bounded near $a_t$, $b_t$, and $x^*$,
\end{itemize}
where $J_S$ is given by
\begin{equation}
J_S=\begin{cases}\begin{array}{ll}
\begin{pmatrix}
  1 & e^{-2n \varphi_t} \\
  0 & 1
\end{pmatrix} &\textrm{ on } (b_t,\infty), \\[3ex]
\begin{pmatrix}
  1 & e^{-2n \tilde{\varphi}_t} \\
  0 & 1
\end{pmatrix} &\textrm{ on } (-\infty,a_t), \\[3ex]
\begin{pmatrix}
  0 & 1 \\
  -1 & 0 \\
\end{pmatrix} &\textrm{ on } (a_t,b_t),\\[3ex]
\begin{pmatrix}
  1 & 0 \\
  e^{2n \varphi_t(z)} & 1 \\
\end{pmatrix} &\textrm{ on } C_1 \cup C_2,
\end{array}
\end{cases}\label{3-sprongvS}
\end{equation}

The jump matrices on $\mathbb R \setminus (a_t,b_t)$ and on
$C_1 \cup C_2$ tend to the identity matrix as
$n \to \infty$ and $t \to 1$. Ignoring these jumps, we
find the parametrix for the outside region. Uniform convergence breaks
down in neighborhoods of $a_t$, $b_t$, and $x^*$, so that we will
also need to construct local parametrices near those points.

\subsection{Parametrix away from special points}
The outside parametrix $S^{\infty} = S^{\infty}_t$ solves the
following RH problem
\begin{itemize}
\item[(a)] $S^{\infty}$ is analytic in $\mathbb C\setminus [a_t,b_t]$,
\item[(b)] $S_+^{\infty}=S_-^{\infty}
    \begin{pmatrix} 0 & 1 \\ -1 & 0 \end{pmatrix} $ on
    $(a_t, b_t)$,
\item[(c)] $S^{\infty}(z)=I+O(z^{-1})$ as $z\to \infty$.
\end{itemize}
As in \cite{DKMVZ2} it has the solution
\begin{equation} \label{def-Sinfty}
S^{\infty}_t(z) = \begin{pmatrix}
  \frac{\beta(z)+\beta(z)^{-1}}{2} & \frac{\beta(z)-\beta(z)^{-1}}{2i} \\
  -\frac{\beta(z)-\beta(z)^{-1}}{2i} & \frac{\beta(z)+\beta(z)^{-1}}{2} \\
\end{pmatrix},
    \qquad z \in \mathbb C \setminus [a_t,b_t],
\end{equation}
where
\begin{equation} \label{def-beta}
\beta(z)= \beta_t(z) = \left(\frac{z-b_t}{z-a_t}\right)^{1/4},
    \qquad z \in \mathbb C \setminus [a_t,b_t].
\end{equation}
Note that $S^{\infty}_t$ depends on $t$.

\subsection{Parametrix at edge points}
At edge points $a_t$ and $b_t$ the density $\psi_t$ vanishes like
a square root. This allows the construction of local parametrices
near $a_t$ and $b_t$ with the use of Airy functions. We will not
give the details, see \cite{Deift,DKMVZ1,DKMVZ2}.

\subsection{Parametrix at critical point}
In a neighborhood $ \Delta = \{ z \in \mathbb C \mid |x^*-z| < \delta \}$
of $x^*$, we want to
construct a parametrix $P$ with the following properties
\begin{itemize}
\item[(a)] $P$ is analytic in $\Delta\setminus (\mathbb R\cup
C_1\cup C_2)$,
\item[(b)] ${P}_+={P}_- J_S(z)$ on
$(\mathbb R\cup C_1\cup C_2) \cap \Delta$,
\item[(c)] $P(z) =  (I + O(n^{-1/3})) S_t^{\infty}(z)$
as $n\to \infty$, $t \to 1$,  uniformly
for $z \in \partial \Delta$,
\item[(d)] $P(z)$ remains bounded for $z$ near $x^*$.
\end{itemize}

We seek a parametrix $P$ near $x^*$ in the form
\begin{equation}\label{5-constructieP}
    P(z) = E_t(z) M\left(n^{1/3} f(z); n^{2/3} s_t(z), n \theta_t\right)
    \end{equation}
where $E_t$ is analytic in $\Delta$,
$f$ is a conformal map from $\Delta$
to a neighborhood of $0$, $s_t$ is analytic
in $\Delta$, and $\theta_t$ is a real constant.
Recall that $M$ is given by (\ref{def-M}).

In view of the jump properties of $M$, we seek
$f$, $s_t$, and $\theta_t$ so that
for $z \in \Delta$,
\begin{equation} \label{mappings1}
    i\left(\frac{4}{3} f(z)^3 + s_t(z) f(z) - \theta_t \right)
    = \begin{cases} \begin{array}{rl}
     \varphi_t(z) & \qquad \mbox{ if } \Im z > 0, \\
     - \varphi_t(z) & \qquad \mbox{ if } \Im z < 0.
     \end{array} \end{cases}
\end{equation}
Since ${\varphi_t}_+ = - {\varphi_t}_-$,
the right-hand side of (\ref{mappings1}) does indeed
define an analytic function in $\Delta$.

The conformal map $f$ will not depend on $t$. Recall
that $q_V$ has a zero of order four at $x^*$, and that
it is negative on $(a,b) \setminus \{x^*\}$. Thus
$\int_{x^*}^z (-q_V(y))^{1/2} dy$ is analytic in $\Delta$
and has a zero of order three at $x^*$. We may assume there
are no other zeros in $\Delta$. Then we can take
a third root and define
\begin{equation} \label{def-fz}
    f(z) = \left[\frac{3}{4}
    \int_{x^*}^z (-q_V(y))^{1/2} dy \right]^{1/3},
    \qquad z \in \Delta.
\end{equation}
Then $f$ is analytic in $\Delta$ with
 \begin{align} \nonumber
    f(z) & = \left( \frac{\pi \psi_V''(x^*)}{8}\right)^{1/3}
    (z-x^*) + O\left((z-x^*)^2\right)
    \\ & = c^{1/3} (z-x^*) + O\left((z-x^*)^2\right),
    \qquad \mbox{ as } z \to x^*
    \label{f-near-xstar}
\end{align}
where $c$ is given by (\ref{main2}).
Taking smaller $\Delta$ if necessary, we then have that $f$ is
indeed a conformal map on $\Delta$. Note that $f(x^*) = 0$ and
that $f$ is real and positive on $(x^*, x^*+\delta)$ and real and
negative on $(x^*-\delta, x^*)$. We still have some freedom in
opening the lenses. We take $C_1$ and $C_2$ in $\Delta$ so that
$f$ maps them to the rays where $M$ has its jumps. That is, $C_1
\cap \Delta$ is mapped into $\arg \zeta = \pi/6$ and $\arg \zeta =
5 \pi/6$, and $C_2 \cap \Delta$ is mapped into $\arg \zeta = -
\pi/6$ and $\arg \zeta = - 5\pi/6$.

Having $f$ with $f(x^*) = 0$, we
take $z = x^*$ in (\ref{mappings1}) and we see that
we should take
\begin{equation} \label{def-theta1}
    \theta_t = i{\varphi_t}_+(x^*) = -i {\varphi_t}_-(x^*).
\end{equation}
Then $\theta_t$ is real and it is also given by
\[ \theta_t = - \int_{b_t}^{x^*} (-q_t(y))^{1/2} dy \]
and so
\begin{equation} \label{def-theta2}
    \theta_t \mp i \varphi_t(z) =
    \int_{x^*}^z (-q_t(y))^{1/2} dy,
    \qquad \mbox{ for } \pm \Im z > 0.
\end{equation}

Having $f$ and $\theta_t$ we finally take
$s_t(z)$ so that (\ref{mappings1}) holds, that is,
\begin{align}
    s_t(z) f(z) & = \nonumber
    -\frac{4}{3} f(z)^3 + \theta_t \mp i \varphi_t(z) \\
    & = \int_{x^*}^z \left( (-q_t(y))^{1/2} - (-q_V(y))^{1/2}\right) dy
    \label{def-st}
    \end{align}
where for the last line we used (\ref{def-fz}) and (\ref{def-theta2}).
Since the right-hand side of (\ref{def-st}) is analytic in $\Delta$
and vanishes for $z = x^*$ we can divide by $f(z)$
(which has a simple zero at $z=x^*$) and obtain
an analytic function $s_t$ in $\Delta$.
Then $s_t(z)$ is real for real $z$, and
$s_1(z) \equiv 0$.

So the above construction yields analytic functions
$f$, $s_t$ and a constant $\theta_t$ so that
(\ref{mappings1}) holds. Then we can define the
parametrix $P$ by (\ref{5-constructieP}), provided
that $n^{2/3}s_t(z)$ for $z \in \Delta$ stays away
from the poles of the Hastings-McLeod solution of
Painlev\'e II, since for $s \in \mathcal P$ we have
that $M(\zeta;s, \theta)$ is not defined.
For given values of $n$ and $t$, we can take $\Delta$
small enough so that this is indeed the case.
However,  we want to let $n \to \infty$, $t \to 1$
so that $n^{2/3}(t-1) \to L$, and work with a
neighborhood $\Delta$ that is independent of $n$
and $t$, although it may depend on $L$.

Note that by (\ref{def-st}) and (\ref{qt3})
\begin{align}
    s_t(z) f(z) & =
     \pi \int_{x^*}^z (\psi_t(y) - \psi_V(y)) dy.
    \label{def-st2}
\end{align}
Because of (\ref{diff-tpsi}) we have
\[  \psi_t(y) - \psi_V(y)
     = (t-1) (w_{S_V}(y) - \psi_1(y)) + o(t-1),
\]
as $t \to 1$, uniformly for $y$ in a neighborhood of $x^*$.
Using this in (\ref{def-st2}) we get that
\begin{align} \nonumber
    s_t(z) \frac{f(z)}{z-x^*}
    & = \pi (t-1) \frac{1}{z-x^*} \int_{x^*}^z
        (w_{S_V}(y)-\psi_1(y)) dy
        + o(t-1) \\
    & = \pi (t-1) w_{S_V}(x^*) +
        (t-1) O(z-x^*) + o(t-1),
        \label{def-st3}
\end{align}
where $o(t-1)$ is uniformly in $z$ as $t \to 1$,
and $O(z-x^*)$ is uniformly in $t$ as $z \to x^*$.
By (\ref{f-near-xstar}) and (\ref{def-st3})
we then also have
\begin{equation}
    s_t(z) =\pi c^{1/3} (t-1) w_{S_V}(x^*) +
        (t-1) O(z-x^*) + o(t-1),
        \label{def-st4}
\end{equation}
where again $o(t-1)$ is uniformly in $z$ as $t \to 1$,
and $O(z-x^*)$ is uniformly in $t$ as $z \to x^*$.
Then if $n^{2/3} (t-1) \to L$ we get
\begin{align} \nonumber
    n^{2/3} s_t(z) & =
        \pi c^{-1/3} L w_{S_V}(x^*)
            + O(z-x^*) + o(1) \\
            & = \label{def-st5}
            s + O(z-x^*) + o(1),
\end{align}
where we used the definition (\ref{main3}) of
the constant $s$.
Since there are no poles on the real line, we
can find a neighborhood $\Delta$ of $x^*$ such that
$n^{2/3} s_t(z) \not\in \mathcal P$ for all
$z \in \Delta$ if $n$ is large enough
and $n^{2/3}(t-1) \to L$.
Note that $\Delta$ depends on $L$,
but not on $n$ and $t$.
Then
\[ M(n^{1/3} f(z); n^{2/3} s_t(z), n\theta_t) \]
is well-defined for $z$ in
a fixed neighborhood $\Delta$ of $x^*$.

Finally, we define $E_t$ in such a way that the matching
condition at $\partial \Delta$ is satisfied. We do this
by defining
\begin{equation} \label{defEt}
    E_t(z)=
 \begin{cases}\begin{array}{ll}S_t^{\infty}(z) & \textrm{ for } \Im z>0 ,\\
    S_t^{\infty}(z)
    \begin{pmatrix}
  0 & 1 \\
  -1 & 0 \end{pmatrix} & \textrm{ for }  \Im z\leq 0 ,
\end{array}\end{cases}
\end{equation}
so that $E_t(z)$ is analytic near $x^*$
by the jump property (\ref{RHPSinftyb}) of $S_t^{\infty}$.
Because of the asymptotic behavior of $M(\zeta;s, \theta)$
in (\ref{RHPMc1}) and (\ref{RHPMc2}) (which is valid uniformly for $s$ away from the poles
and for $\theta \in \mathbb R$)
we have the matching condition
\[ P(z)= S_t^{\infty}(z) (I+O(n^{-1/3}))
    = (I + O(n^{-1/3})) S_t^{\infty}(z), \]
uniformly for $ z\in\partial \Delta$.
This completes the construction of the
parametrix $P$ in the neighborhood of $x^*$.

\subsection{Third Transformation}

\begin{figure}[h]
\begin{center}
\includegraphics[scale=0.6]{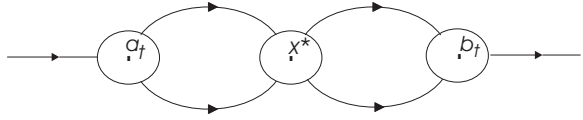}
\end{center}
\caption{The contour $\Gamma$ after the third and final
    transformation.}
\label{figure-contourR}
\end{figure}

We set
\begin{equation}\label{5-RSP}
R(z)= \begin{cases} \begin{array}{ll}
    S(z)P^{-1}(z) & \textrm{ for $z$ in disks around $x^*$, $a$, and $b$}, \\
    S(z)(S_t^{\infty})^{-1}(z) & \textrm{ for $z$ outside the disks}.
    \end{array}    \end{cases}
\end{equation}
Since $S$ and $P$ have the same jumps inside each of the disks,
and $S$ and $S_t^{\infty}$ have the same jumps on $[a_t,b_t]$,
$R$ has jumps on a contour $\Gamma$ as shown in
Figure \ref{figure-contourR}.

$R$ solves the following RH problem:
\begin{itemize}
\item[\rm (a)] $R$ is analytic in $\mathbb C\setminus \Gamma,$
\item[\rm (b)] $R_+(z)=R_-(z)J_R(z)$ as $z\in \Gamma $,
\item[\rm (c)] $R(z)=I+O(z^{-1})$ as $z\to \infty$,
\item[\rm (d)] $R$ is bounded,
\end{itemize}
for certain jump matrices $J_R$, which as
$n \to \infty$, $t \to 1$ so that
$n^{2/3}(t-1) \to L$, satisfy
\begin{equation*}
J_R(z)=\begin{cases}\begin{array}{ll}
    I + O(n^{-1})&\textrm{ for $z$ in circles around $a$ and $b$} \\
    I + O(n^{-1/3})&\textrm{ for } z\in \partial \Delta,\\
    I + O(e^{-\gamma n})& \textrm{ for some fixed }\gamma>0 \textrm{
elsewhere on $\Gamma$.}
\end{array}\end{cases}
\end{equation*}

As in \cite{Deift,DKMVZ1} it now follows that
\begin{equation} \label{R-asymp}
\|R(z)-I\|_\infty=O(n^{-1/3})
\end{equation}
uniformly for $z \in \mathbb C \setminus \Gamma$.

\section{Proof of Theorem \ref{theorem-1-maintheorem}}

Now we are ready for the proof of Theorem \ref{theorem-1-maintheorem}.
We start with the expression (\ref{3-KnNY}) for the kernel $K_{n,N}$.
After the transformation (\ref{3-TY}), we find that
for $x, y \in (a_t,b_t)$,
\begin{equation}\label{6-kernT}
K_{n,N}(x,y)=\frac{1}{2\pi i(x-y)}
\begin{pmatrix}
  0 & e^{-n{\varphi_t}_+(y)} \end{pmatrix}
    T_+^{-1}(y)T_+(x)\begin{pmatrix}
    e^{-n{\varphi_t}_+(x)} \\
    0 \end{pmatrix}.
\end{equation}
Using formula (\ref{3-ST}) for $S$ in the upper parts
of the lenses, we get for $x,y \in (a_t,b_t)$,
\begin{equation} \label{6-kernS}
K_{n,N}(x,y) = \frac{1}{2\pi i(x-y)}
\begin{pmatrix}
-e^{n {\varphi_t}_+(y)} & e^{n {\varphi_t}_+(y)}
\end{pmatrix} S_+^{-1}(y)
S_+(x) \begin{pmatrix}
e^{-n {\varphi_t}_+(x)} \\
e^{n {\varphi_t}_+(x)}
\end{pmatrix}
\end{equation}
Assume that $x$ and $y$ are inside the disk $\Delta$
around $x^*$. Then by (\ref{5-RSP}), (\ref{5-constructieP}),
(\ref{def-M}), and (\ref{mappings1}) we get
\[ S_+(x) = R(x) E_t(x) e^{i n \theta_t \sigma_3}
    \Psi_+(n^{1/3} f(x); n^{2/3} s_t(x)) e^{n {\varphi_t}_+(x) \sigma_3},
     \]
and so
\begin{equation} \label{Sformula1}
S_+(x) \begin{pmatrix}
e^{-n {\varphi_t}_+(x)} \\
e^{n {\varphi_t}_+(x)}
\end{pmatrix}
    = R(x) E_t(x) e^{in \theta_t \sigma_3}
    \Psi(n^{1/3} f(x); n^{2/3} s_t(x))
    \begin{pmatrix} 1 \\ 1 \end{pmatrix}.
    \end{equation}
Similarly
\begin{align} \nonumber
    \lefteqn{
    \begin{pmatrix}
-e^{n {\varphi_t}_+(y)} & e^{n {\varphi_t}_+(y)}
\end{pmatrix}  S_+^{-1}(y)} \\ \label{Sformula2}
& = \begin{pmatrix} -1 & 1 \end{pmatrix}
    \Psi^{-1}(n^{1/3} f(y); n^{2/3} s_t(y))
    e^{-in \theta_t \sigma_3}
    E_t^{-1}(y) R^{-1}(y).
    \end{align}

Now we fix $u$ and $v$ and  take
\begin{equation} \label{xy-in-uv}
    x = x^* + \frac{u}{(cn)^{1/3}} \quad \mbox{ and } \quad
    y = x^* + \frac{v}{(cn)^{1/3}},
\end{equation}
so that for $n$ large enough,
$x$ and $y$ are inside the disk around $x^*$, so that
(\ref{6-kernS}), (\ref{Sformula1}), and (\ref{Sformula2}) hold.
Then it follows from (\ref{f-near-xstar}) and
(\ref{xy-in-uv}) that
\begin{equation} \label{f-limits}
  n^{1/3} f(x) \to u, \quad \mbox{ and }
    \quad
  n^{1/3} f(y) \to v \qquad \mbox{ as } n \to \infty.
  \end{equation}
From (\ref{def-st5}) and (\ref{xy-in-uv}) we
get
\begin{equation} \label{st-limits}
 n^{2/3} s_t(x) \to s, \quad \mbox{ and } \quad
 n^{2/3} s_t(y) \to s
 \end{equation}
as $n \to \infty$, $t \to 1$ such that $n^{2/3}(t-1) \to L$.
Furthermore, from (\ref{R-asymp}), (\ref{xy-in-uv}), and the fact that $R$ is
analytic near $x^*$, we get
\begin{equation} \label{R-limits}
 R^{-1}(y) R(x) = I + O\left(\frac{x-y}{n^{1/3}}\right)
    = I + O \left(\frac{u-v}{n^{2/3}} \right).
\end{equation}
From (\ref{def-Sinfty}), (\ref{defEt}), and (\ref{xy-in-uv})
we easily get
\begin{equation} \label{Et-limits}
 E_t^{-1}(y) E_t(x) = I + O\left(\frac{u-v}{n^{1/3}}\right)
\end{equation}
as $n \to \infty$. The constants implied by the $O$-symbols in
(\ref{R-limits}) and (\ref{Et-limits}) are independent of
$u$ and $v$, when $u$ and $v$ are restricted to a compact subset of $\mathbb R$.
Combining (\ref{R-limits}) and (\ref{Et-limits}) we get that
\begin{equation} \label{RandEt-limits}
e^{-i n\theta_t \sigma_3}E_t^{-1}(y)R^{-1}(y) R(x) E_t(x)
e^{i n\theta_t \sigma_3} = I + O\left(\frac{u-v}{n^{1/3}}\right),
\end{equation}
since $\theta_t$ is real.

Then multiplying (\ref{Sformula2}) and (\ref{Sformula1})
and letting $n \to \infty$, $t \to 1$ such that
$n^{2/3} (t-1) \to L$, we get by using (\ref{6-kernS}), (\ref{xy-in-uv}), (\ref{f-limits}),
(\ref{st-limits}), and (\ref{RandEt-limits}) that
\begin{eqnarray} \nonumber
\lefteqn{ \lim_{n \to \infty}
\frac{1}{(cn)^{1/3}} K_{n,N}
\left(x^* + \frac{u}{(cn)^{1/3}}, x^* + \frac{v}{(cn)^{1/3}}\right)} \\
& & = \nonumber
\frac{1}{2\pi i(u-v)}
\lim_{n \to \infty} \begin{pmatrix}
-e^{n {\varphi_t}_+(y)} & e^{n {\varphi_t}_+(y)}
\end{pmatrix}  S_+^{-1}(y)
S_+(x) \begin{pmatrix}
e^{-n {\varphi_t}_+(x)} \\
e^{n {\varphi_t}_+(x)}
\end{pmatrix} \\
& & =
  \label{Sformula3} \frac{1}{2\pi i(u-v)}
\begin{pmatrix} -1 & 1 \end{pmatrix}
 \Psi^{-1}(v;s) \Psi(u;s)
\begin{pmatrix} 1 \\ 1 \end{pmatrix},
\end{eqnarray}
uniformly for $u$ and $v$ in compact subsets of $\mathbb R$.

By (\ref{1-asymptotiekphi}), (\ref{Psij}) and (\ref{def-Psi}) we have that
\[ \begin{pmatrix} \Phi_1(\zeta;s) \\ \Phi_2(\zeta;s) \end{pmatrix} =
   \Psi(\zeta;s) \begin{pmatrix} 1 \\ 0 \end{pmatrix}
   \qquad \mbox{for } \zeta \in S_2 \cup S_3. \]
In view of the jump (\ref{RHPPsib1}) satisfied by $\Psi(\zeta;s)$
on $\Sigma_1$ we have that
\begin{equation} \label{Phi-in-Psi1}
    \begin{pmatrix}
    \Phi_1(\zeta; s) \\
    \Phi_2(\zeta; s) \end{pmatrix}
    = \Psi(\zeta;s) \begin{pmatrix} 1 \\ 1
    \end{pmatrix} \qquad \mbox{for } \zeta \in S_1 \cup S_4,
    \end{equation}
Since $\det \Psi \equiv 1$, we also get (after
simple calculation)
\begin{equation} \label{Phi-in-Psi2}
    \begin{pmatrix} -1 & 1 \end{pmatrix}
    \Psi^{-1}(\zeta;s) =
    \begin{pmatrix} - \Phi_2(\zeta;s) & \Phi_1(\zeta;s)
    \end{pmatrix}
    \quad \mbox{for } \zeta \in S_1 \cup S_4.
\end{equation}
Note that $S_1 \cup S_4$ includes the full real line, so
that we can take $\zeta = u$ in (\ref{Phi-in-Psi1})
and $\zeta=v$ in (\ref{Phi-in-Psi2}) which we use
in (\ref{Sformula3}) to obtain (\ref{main1}).
This completes the proof of Theorem \ref{theorem-1-maintheorem}.

\begin{remark}\label{remark-5-vereenvoudiging}
The above proof of Theorem \ref{theorem-1-maintheorem} was given
under the assumption that $S_V$ consists of one interval, and that
there are no other singular points except for $x^*$.
Here singular point refers to the classification of \cite{DKMVZ2}
according to which there are three types of non-regular behavior for
a real-analytic external field $V$, see also \cite{KM}.
The singular points of type I
are points in $\mathbb R \setminus S_V$ where equality in
the variational inequality (\ref{variational-mut2}) with
$t=1$ holds. Singular points of type II are interior
points of $S_V$ where the density $\psi_V$ vanishes, and
singular points of type III are edge points of $S_V$ where
$\psi_V$ vanishes to higher order than a square root.
Here we indicate briefly the modifications that have to be
made if these assumptions are not satisfied.

If $S_V$ consists of more than one interval, then the main
complication is that the construction of the outside parametrix
$S^{\infty}_t$ is more complicated, since it uses $\Theta$-functions
as in \cite[Lemma 4.3]{DKMVZ2}.
It should be noted that $S^{\infty}_t$ will also
depend on $n$. As a result it will follow that
$E_t$ as defined in (\ref{defEt}) also depends on $n$.
However this will not effect the asymptotic behavior
(\ref{Et-limits}) as $n \to \infty$, so that the above
proof goes through.

In case there are other singular points, we have to
construct special local parametrices around each of them.
For our purposes, we only need to know the existence
of those parametrices, which is established in
\cite[Section 5]{DKMVZ2}.
\end{remark}

\section*{Acknowledgements}
We thank Maarten Vanlessen for useful remarks.

The authors are supported by FWO research projects G.0176.02
and G.0455.04.
The second author is also supported by K.U.Leuven research grant OT/04/24,
by  INTAS Research Network NeCCA 03-51-6637,
by NATO Collaborative Linkage Grant PST.CLG.979738,
by grant BFM2001-3878-C02-02 of the Ministry of
Science and Technology of Spain and
by the European Science Foundation Program
Methods of Integrable Systems, Geometry, Applied Mathematics
(MISGAM)
and the European Network in Geometry,
Mathematical Physics and Applications (ENIGMA).

\obeylines
\texttt{
Tom Claeys
Department of Mathematics
Katholieke Universiteit Leuven
Celestijnenlaan 200B
B-3001 Leuven, BELGIUM
E-mail: tom.claeys@wis.kuleuven.ac.be
\bigskip

Arno B.J. Kuijlaars
Department of Mathematics
Katholieke Universiteit Leuven
Celestijnenlaan 200B
B-3001 Leuven, BELGIUM
E-mail: arno@wis.kuleuven.ac.be
}
\end{document}